\documentclass[a4paper,11pt]{article}
\pdfoutput=1 

\usepackage{jcappub} 

\usepackage[T1]{fontenc} 
\usepackage{caption}
\usepackage{subcaption}
\usepackage{float}
\usepackage{multirow}
\usepackage{array}
\usepackage{amsmath}
\usepackage{makecell}
\usepackage{xcolor}
\usepackage{slashbox}
\usepackage{ctable}
\usepackage{boldline}

\graphicspath{
	  {./} 
    {./figures/}    
}

\newcolumntype{C}[1]{>{\centering\let\newline\\\arraybackslash\hspace{0pt}}m{#1}}
\newcommand{\thickhline}{%
    \noalign {\ifnum 0=`}\fi \hrule height 1pt
    \futurelet \reserved@a \@xhline
}
\newcolumntype{"}{@{\hskip\tabcolsep\vrule width 1pt\hskip\tabcolsep}}

\newcommand{\ie}{{\it i.e.}}

\newcommand{\eg}{{\it e.g.}}

\newcommand{\etc}{{\it etc.}}
\newcommand{\eq}{Eq.}

\newcommand{\Ref}{Ref.}
\newcommand{\Refs}{Refs.}
\newcommand{\Sec}{Section}

\newcommand{\Tab}{Tab.}

\newcommand{\rmd}{{\rm d}}

\newcommand{\equ}[1]{\eq~(\ref{eq:#1})}
\newcommand{\sect}[1]{\Sec~\ref{sec:#1}}
\newcommand{\figu}[1]{Fig.~\ref{fig:#1}}

\definecolor{deepmagenta}{rgb}{0.8, 0.0, 0.8}

\title{\boldmath Improved photomeson model for interactions of cosmic ray nuclei}

\author[a]{L. Morejon\footnote{Corresponding author.}}
\author[a,b]{A. Fedynitch}
\author[a,c,d]{D. Boncioli}
\author[a]{D. Biehl}
\author[a]{\\W. Winter}

\affiliation[a]{Deutsches Elektronen-Synchrotron (DESY), Platanenallee 6, 15738 Zeuthen, Germany}
\affiliation[b]{Dept. of Physics, University of Alberta, Edmonton, Alberta, Canada T6G 2E1}
\affiliation[c]{Gran Sasso Science Institute (GSSI), Viale Francesco Crispi 7, 67100 L'Aquila, Italy}
\affiliation[d]{INFN, Laboratori Nazionali del Gran Sasso (LNGS), 67100 Assergi, L'Aquila, Italy}
\emailAdd{leonel.morejon@desy.de}
\emailAdd{anatoli.fedynitch@desy.de}
\emailAdd{denise.boncioli@gssi.it}
\emailAdd{daniel.biehl@desy.de}
\emailAdd{walter.winter@desy.de}

\abstract{
	Photon-hadronic interactions are important for the sources and the transport of Ultra-High Energy Cosmic Rays (UHECRs). Current state-of-the-art cosmic ray transport simulations handle nuclear disintegration at energies of the Giant Dipole Resonance at a more sophisticated level, as well as the photohadronic interactions of nucleons in the high-energy regime above the pion production threshold. However, the interactions of nuclei above the pion production threshold are commonly modeled by treating the nucleus as a superposition of free nucleons -- ignoring the effect of the nuclear medium. We construct an improved, inclusive model for the photomeson regime for nuclei with $A \leq 56$ by employing more accurate, data-driven parametrizations of the interaction cross section, the fragmentation of the primary nucleus and the inclusive pion production cross section that directly affects the production of astrophysical neutrinos. We apply our results to two multi-messenger scenarios (Tidal Disruption Events and Gamma-Ray Bursts) in which photonuclear interactions in the photomeson regime are the dominant cooling process for the highest energy cosmic rays. While we find moderate changes to the mass composition of UHECRs, the astrophysical neutrino fluxes exhibit a significant (factor of a few) reduction compared to the na\"ive superposition of free nucleons for sources of UHECR nuclei with a populated cascade. The numerical code implementing the model has been made publicly available, which facilitates the integration of our results in similar frameworks.
}

\begin{document}
\maketitle

\section{Introduction}
	\label{sec:intro}

The mass composition measurements of Ultra-High Energy Cosmic Rays ($E > 10^{18}$ eV) from the 
Pierre Auger Observatory and the Telescope Array are clearly compatible to a mixed composition 
\cite{deSouza:2017}, supporting the presence of nuclei heavier than protons in astrophysical 
accelerators, such as Active Galactic Nuclei~\cite{PhysRevLett.69.2885}, Gamma Ray 
Bursts~\cite{waxman1995cosmological} or Tidal Disruption 
Events~\cite{Farrar:2014yla,10.1093/mnras/stw3337}. Interactions of UHECR occur near the 
acceleration site for high in-source photon densities and during the transport between source and 
Earth with the cosmic microwave background (CMB), the infrared and the optical photons. We refer to 
these interactions as photohadronic interactions of nuclei.

Photohadronic interactions occur at different energy scales~\cite{Rachen:1996ph}, where the 
energy scale typically refers to the photon energy in the nucleus' rest frame $\epsilon_r$.  At 
low energies $\epsilon_r \gtrsim 1$ MeV hadronic processes can not occur, hence particles are 
produced via electromagnetic (Bethe-Heitler)  
electron-positron pair production. At energies comparable to the nuclear binding energy
$\epsilon_r \gtrsim 8$ MeV, a resonant motion of the nucleons leads to the disintegration of the 
nucleus through the excitation of the Giant Dipole Resonance (GDR). Models that parametrize this 
process for astrophysical applications are based on tables by (for example) Puget-Stecker-Bredekamp 
(PSB) \cite{Puget:1976nz,Stecker:1998ib}, or are computed from running nuclear event generators, 
such as {\sc Talys}~\cite{talys18}, {\sc GEANT 4}~\cite{Agostinelli:2002hh} or 
{\sc Fluka}~\cite{fluka}. 
Beyond $\epsilon_r \gtrsim 140 $ MeV  (hadronic scale) hadronic processes start, leading to 
the production of baryonic resonances that emit pions and photons when they decay. The lowest mass 
resonances, such as the $\Delta$-resonance, are produced in the $s$-channel for photon projectiles 
and are thus more efficient in dissipating the available energy into secondary particles compared
to proton-proton collisions where they are produced from t-channel processes at threshold. At higher 
energies $\epsilon_r \gtrsim 1$ GeV, the hadronic mass scale, the photon interacts hadron-like as a 
virtual vector meson. We refer to the latter two energy scales ($\epsilon_r \gtrsim 140 $ MeV) as 
``photomeson'' regime.

The physics of photomeson production in astrophysical environments has been studied extensively
for nucleons (protons and neutrons) as early as \cite{Greisen:1966jv,Zatsepin:1966jv} and later in 
\Refs~\cite{Berezinsky:1992ds,Berezinsky:1993im,Mannheim:1994sv,Gaisser:1994yf,Rachen:1996ph,Mucke:1998ms}. 
The {\sc Sophia}~\cite{Mucke:1999yb} Monte Carlo code is tailored to simulate photon-nucleon 
interactions around the threshold, including the most important baryonic resonances and direct 
production ($t$-channel) processes. The model at higher energies is somewhat simpler, but 
sufficiently precise for modeling multi-pion production in the scope of astrophysical applications. 
Different parametrizations using {\sc Sophia} for semi-analytical and numerical transport equation 
solvers have been developed in \Refs~\cite{Kelner:2008ke,Hummer:2010vx,Biehl2018a}. Nuclei, however,
are treated in the superposition approach, in which photons interact with one nucleon and the 
remainder is left intact. For this single photon-nucleon interaction, the differential cross 
sections for secondary particles are sampled from {\sc Sophia}, whereas the remnant nucleus simply 
has the mass $A-1$ neglecting momentum transfer or recoil, see \eg\ 
\Refs~\cite{Anchordoqui:2007tn,Murase:2008mr,Batista_2016,boncioli2016nuclear,Aloisio:2017iyh}.
The effective number of target nucleons seen by the photon projectile is reflected in the ``mass 
scaling'' of the cross section $A_{\rm eff} = \sigma_{A\gamma}/\sigma_{p\gamma} \sim A^\alpha$. The
power alpha lies between two extreme values: $1$ when all nucleons are seen by the photon as targets, 
and $2/3$ when only nucleons at the surface are active targets and those inside are screened.
\Refs~\cite{Kampert:2012fi,Batista_2016} make a global scaling assumption for the entire photomeson
energy range. In reality, different nuclear scaling powers $\alpha$ are appropriate for the total
interaction and the pion production cross sections, with different values for energies close to the
threshold and for very high energies.

If photomeson production is the dominant cooling process in a UHECR source, the disintegration/
fragmentation of the UHECR will be described by the interactions above the $\Delta$-resonance 
threshold. For intergalactic UHECR propagation, the giant dipole resonance (GDR) dominates the 
photohadronic processes because the energy and high density of the CMB photons constitutes the 
dominant cooling process. This happens at UHECR energies below the photomeson threshold, hence 
lowering the importance of the photomeson model for propagation calculations. Under certain 
circumstances the photomeson regime can dominate the photohadronic interactions in astrophysical 
sources, if the UHECR does not find interaction partners (photons) to match the GDR energy. 
Known examples of this are Gamma-Ray Bursts, with a minimal photon energy cutoff~\cite{Biehl2018a} 
which may come from synchrotron self-absorption~\cite{Murase:2008mr}, and jetted Tidal Disruption 
Events involving massive stars~\cite{Biehl2018b} with a spectral photon index such that photon 
densities of lower energy photons are smaller than usual, making photomeson production the  dominant
interaction for UHECRs of at the highest energies. We will use these as astrophysical  examples in
our discussion of the impact of the photomeson model.

Apart from the primary UHECRs, the neutrino production is always related to the photomeson regime: 
if the neutrinos are mostly produced off nuclei, corrections to the photomeson model are expected 
to have an impact, whereas if the neutrinos are produced off nucleons, the physics is well 
described by, for instance, {\sc Sophia}. The former situation mostly occurs in sources with low
radiation fields, which means that the primary nucleus barely disintegrates, whereas the latter
situation is typical for sources with strong radiation fields in which nucleons emerging from
a nuclear cascade dominate the neutrino production~\cite{Biehl2018a}.

In this paper we will introduce an improved  model for the photomeson production and compare it to
the widespread ``Single Particle Model'' (SPM). The improvements involve modifications relevant for
the astrophysical processes: improved cross section descriptions driven by data, and  empirical
approaches to describe the fragmentation of the nucleus. The impact of these  modifications is
illustrated at astrophysical simulations of Gamma-Ray Bursts (GRBs) and Tidal Disruption  Events
(TDEs). The paper is organized as follows: \Sec~\ref{sec:photon_interactions} discusses the 
physical differences between photon interactions with nucleons and with nuclei, and introduces the 
quantities and nomenclature related to multi-messenger astroparticle physics simulations.
\Sec~\ref{sec:baseline_model} presents the ingredients of the new photomeson model, and 
\Sec~\ref{sec:source_scenarios} illustrates the impact of the model in astrophysical source
simulations  in which the photomeson model is relevant.

\section{Astrophysical photohadronic interactions: \texorpdfstring{$\boldsymbol{p \gamma}$}{p-gamma} 
vs \texorpdfstring{$\boldsymbol{A \gamma}$}{A-gamma}} 
\label{sec:photon_interactions}

A convenient reference frame to discuss photonuclear 
interactions is the nuclear rest frame, in which the photon energy
${\epsilon_r = E\, \varepsilon \,(1 - \cos\theta) / m_j}$ depends on the energy of the relativistic 
nucleus $E$ and the photon energy $\varepsilon$ 
in the observer's or (cosmological) comoving frame. The pitch angle $\theta$ is the angle between
incident photon and nucleus such that $\cos \theta = -1$ represents head-on collisions, and
$\epsilon_r$ is related to the center-of-mass energy by $s=m_j^2+2 m_j \epsilon_r$ where $m_j$ 
is the mass of the nucleus. The photonuclear interaction rate and the interaction cross section
$\sigma$ are related as
\begin{align}
	\label{eq:interaction_rate}
	\Gamma(E) = \int \text{d} \varepsilon \int_{-1}^{+1} &\frac{\text{d}\cos\theta}{2} (1 - \cos 
	\theta) n_\gamma(\varepsilon, \cos\theta) \, \sigma(\epsilon_r),
\end{align}
where $n_\gamma$ is the photon number density and the rate is expressed in units of inverse length.
Depending on the type of source or environment, the photon spectrum 
can extend from sub-eV up to TeV energies, and its shape can contain peaked (thermal) or power-law (non-thermal) components. 

\begin{figure}[t]
	\centering
	\includegraphics[width=.6\textwidth]{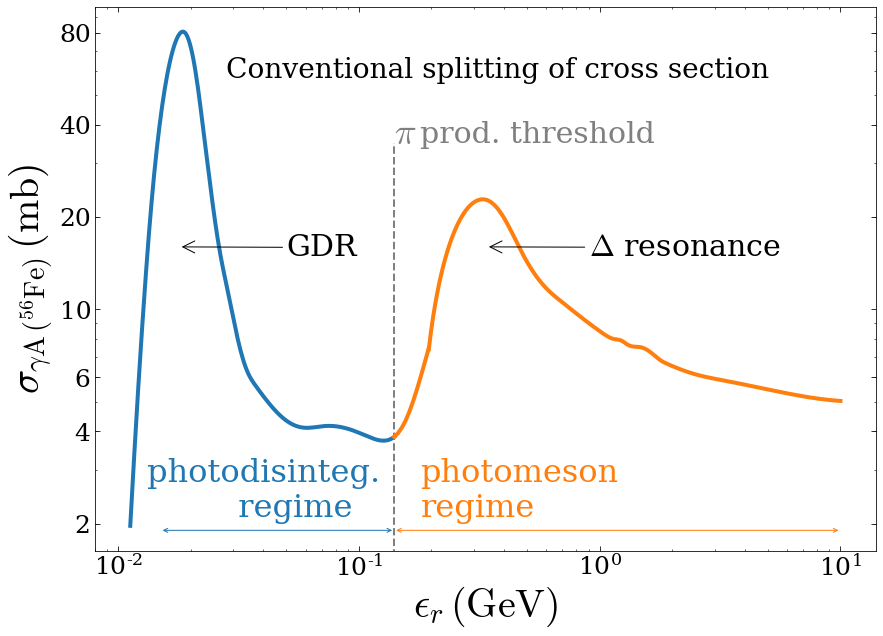}
	\caption{The total inelastic photonuclear cross section for $\rm ^{56}Fe$ as a function of 
	photon energy in the nucleus' rest frame illustrates the general shape for nuclei. The 
	convention of distinguishing two regions based on the photon energy is represented with a 
	change  of the color. The photodisintegration portion (in blue) refers to photon energies $\rm 
	\epsilon_r$ below the photopion production threshold ($\sim 140 \, \mathrm{MeV}$), and the the 
	photomeson portion (in orange) refers to photon energies above the photopion production 
	threshold.}
	\label{fig:cross_sec_tot}
\end{figure}

The shape of a typical photonuclear cross section is illustrated in \figu{cross_sec_tot} for  
$\rm ^{56}Fe$. The energy range is split in a photodisintegration regime, characterized by the
absence of hadron production and negligible momentum transfers or recoils compared to typical cosmic
ray energies (boost conservation); and a photomeson regime, with the possibility for production of
all sorts of particles (predominantly pions) but with significant momentum transfer. A more detailed
discussion about the role of photodisintegration for cosmic ray astrophysics is contained in
\eg{}~\cite{boncioli2016nuclear,Batista:2015mea,AlvesBatista:2019rhs}. 

\begin{figure}[t]
	\centering
	\includegraphics[width=0.75\textwidth]{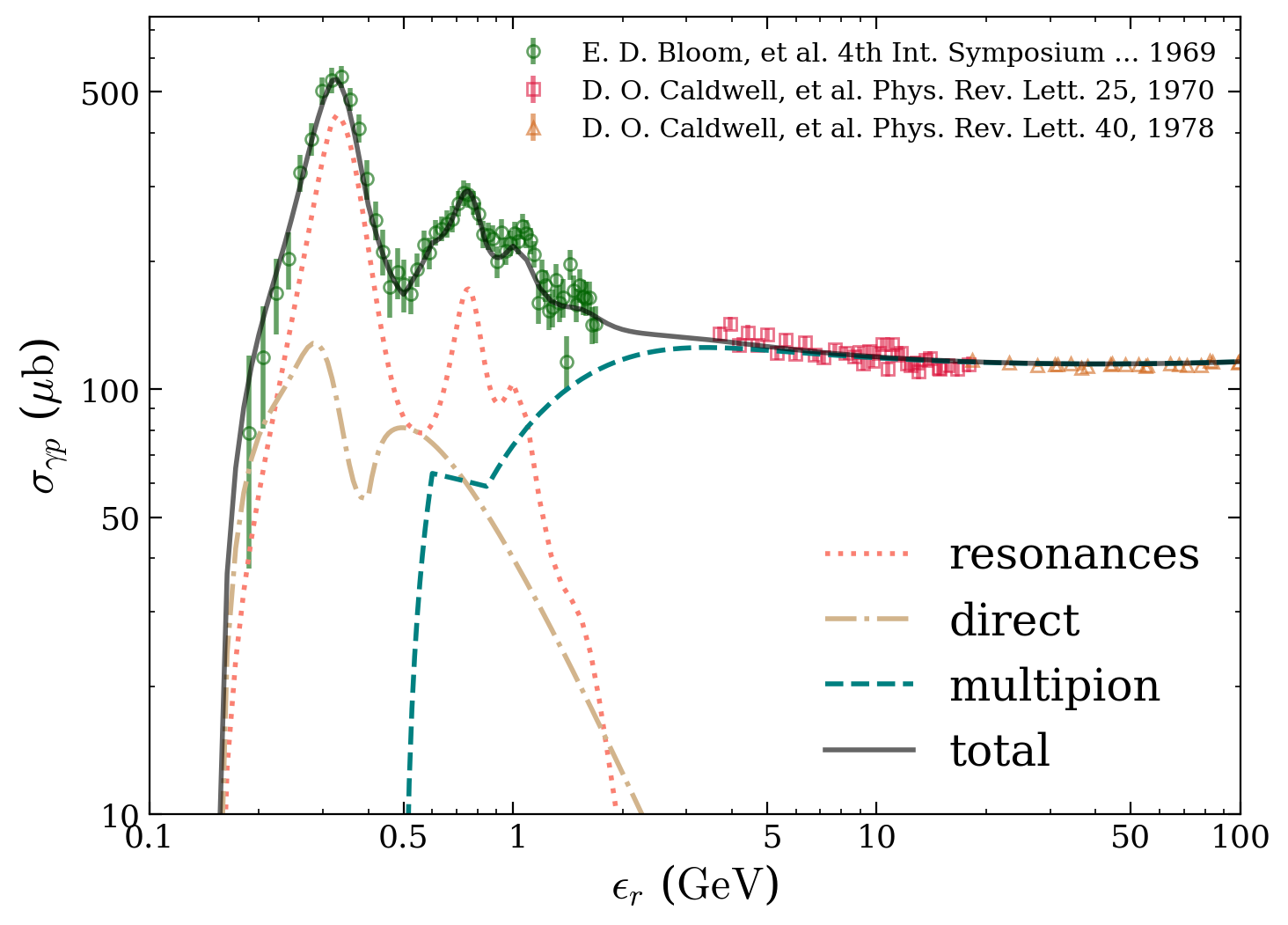}
	\caption{The cross section for inelastic scattering of photons by protons as 
	a function of photon energy in the proton rest frame $\epsilon_r$.
	The curves correspond to the theoretical estimates of different processes. 
	The measurements of the total cross section are from
	\Refs~\cite{Caldwell:1970fw,Caldwell:1978yb,Bloom:1969pn}.}
	\label{fig:pgamma_components}
\end{figure}

It is instructive to first discuss the photomeson production in the absence of collective effects, 
\ie{} in interactions of photons with free nucleons, which has been extensively studied 
experimentally and theoretically~\cite{book:16633,book:springerVol39}. The proton and neutron cross
sections are different below $\approx 140\, \rm MeV$, since for protons Thomson scattering and 
pair-production are possible, whereas only the much weaker magnetic moment scattering occurs for the 
neutron~\cite{Gould:1992ss}. The production of pions at threshold occurs through the excitation of
the lightest baryonic resonance  ($\Delta$) in resonant ($s$-channel) scattering. This process has
no counterpart in $pp$  scattering since there are no known di-baryon resonant states, and hence no
$s$-channel equivalent. Instead, mesons are produced through $t$-channel processes at sufficiently
large momentum transfer. Above the pion threshold more channels are available for the production of
higher mass  resonances and there are small differences between proton and neutron cross sections
(see \figu{pgamma_components}). At high energies above a few GeV, the photon interacts mostly as a
virtual vector meson (see for instance \Ref~\cite{Engel:1994vs}) and all phenomena of hadronic
interactions can occur.

The transport equations contain re-injection terms that represent the production rate of particles 
of species $i$ at energy $E_i$ converting from the particle species $j$ at energy $E_j > E_i$ 
through interactions or decays. The general form for the re-injection rate is
\begin{equation}
	\label{eq:reinjection_rate}
	Q_{ji}(E_i) = \int dE_j \, N_j(E_j) \, \Gamma_j (E_j) \, \frac{\rmd n_{j \to i}}{\rmd E
	_i} (E_j,E_i) \, ,
\end{equation}
where $N_j$ is the density of $j$ particles, $\Gamma_j$ the interaction rate from
\equ{interaction_rate} or a decay rate and 
\begin{equation}
	\label{eq:redist_function}
	\frac{\rmd n_{j \to i}}{\rmd E_i} (E_j,E_i) = \frac{1}{\sigma_j(E_j)} \frac{d
	 \sigma^{\rm incl}_{j \to i}}{\rmd E_i} (E_j,E_i)
\end{equation}
an energy redistribution function. By integrating the inclusive differential cross section 
$d \sigma^{\rm incl}_{j \to i}/{\rmd E_i} \,  (E_j,E_i)$ we define the multiplicity
\begin{equation}
	\label{eq:multiplicity}
	M_{j \to i}(E_j) = \frac{1}{\sigma_j(E_j)} \int \rmd E_i \frac{d \sigma^{\rm incl}_{j
	 \to i}}{\rmd E_i} (E_j,E_i).
\end{equation}
that has the meaning of the average number of particles of species $i$ produced per interaction.

If $i$ and $j$ are both nuclei, the energy redistribution takes a simple form when assuming the 
conservation of boost $E_i/m_i=E_j/m_j$ (or the energy per nucleon)
\begin{equation}
    \label{eq:delta_func_per_nucleus}
    \frac{\rmd \sigma^{\rm incl}_{j \to i}}{\rmd E_i}(E_j, E_i) \approx \sigma_j(
    E_j) M_{j \to i}(E_j)\, \delta \left(E_i - \frac{A_i}{A_j} E_j \right).
\end{equation}
For species $i$ other than the remnant nucleus (such as $\pi^\pm$, $\pi^0$, secondary $\rm p$, $\rm n$ 
and higher mass hadrons), the redistribution function does not have a simple parametrization and has 
to be obtained from Monte Carlo simulations (\eg~{\sc Sophia}).

The simplest extension of the free nucleon interactions to nuclear interactions is referred to 
as {\it Single Particle Model} (SPM) in the cosmic ray astrophysics literature \cite{Kampert:2012fi,
Hummer:2010vx}. It assumes that in the photomeson regime the photon interacts always with one 
nucleon in the nucleus without affecting the rest of the nucleus (quasi-free interaction). 
The final state particles are the products of the $\gamma N$ interaction and one remnant 
nucleus with $A - 1$ nucleons. The inelastic cross section is frequently assumed to scale with
 $A$, \ie{} $\sigma_{A\gamma} = A \sigma_{p\gamma}$, implying that
$\rmd \sigma^{\rm incl}_{j \to i}/\rmd E_i = A_j \, \rmd \sigma^{\rm incl}_{N \to i}/\rmd E_i$.

The simplicity of the SPM allows estimating analytically the relative 
importance of photo-nucleon to photonuclear interactions in astrophysical scenarios.
However the mass scaling of the total cross section is reduced with the increase of photon energy 
$\epsilon_r \gtrsim 1$ GeV. Additionally, nuclear medium effects cause differences in the shape 
of the cross sections. The $\Delta$-resonance in nuclei is broadened and the higher resonances are 
smeared out as a result of the Fermi motion of nucleons and the in-medium properties of the nucleon
resonances and mesons. The production of pions is also strongly affected because of final state 
interactions (FSI) of the photo-nucleon products. Finally, the disintegration of the nucleus is
more important than that considered in the SPM, and multiple fragments can be produced with masses 
different from $A-1$. The photonuclear interaction can lead to photospallation and fission processes
which disintegrate the nucleus into nucleons, deuterons, $\alpha$-particles and larger fragments. The 
fission processes are more important for higher masses than the ones relevant in this paper, but the 
model includes it to allow studies where higher masses might be needed.
In the next section we implement these mechanisms in a photomeson model with the simplicity 
characteristic of the SPM.


\section{A new model: Empirical photomeson Model (EM)}
\label{sec:baseline_model}

We improve the SPM in three main aspects:
\begin{enumerate}
	\item {\bf The absorption cross section}: A ``universal function'' better describes the
	shape of the cross section per nucleon the near the $\Delta$ resonance energy. At higher 
	energies an energy dependent mass scaling exponent is introduced to account for nuclear 
	shadowing effects. Both modifications are motivated and derived from data.
	\item {\bf The pion production cross section}: Pion production is known to be strongly influenced by nuclear medium effects~\cite{Nagl:1991:NPP:2588247,Bloch2007}. The inclusive pion production cross section is derived from data \cite{Krusche2004a} for different nuclei and parameterized with an additional curve and a mass scaling exponent that differs the absorption cross section.
	\item {\bf Nuclear fragmentation}: The photonuclear interactions can result 
	in nuclear breakup through different mechanisms not included in the SPM. We implement two alternatives: an Ablation-Abrasion inspired model and an empirical data-driven parametrization. 
	The latter is found in better agreement with detailed simulations and is chosen as our baseline model.
\end{enumerate}

In the following sub-sections, we assess each of the aspects individually. The model described here
has been made available in a code~\cite{leonel_morejon_2019_2600177} and it can be used for 
reproducing this work, as well as for implementing in other frameworks studying UHECR photomeson
interactions.

\begin{table}[t]
\centering
\begin{tabular}{V{3} c|c V{3} c V{3} c V{3} c V{3}}
\specialrule{.1em}{.05em}{.05em}
\multicolumn{2}{V{3} c V{3}}{\backslashbox{Features}{Models}} & \makecell[c]{Single Particle \\ Model (SPM)} & \makecell[c]{Residual Decay \\ Model (RDM)} & \makecell[c]{Empirical \\ Model (EM)} \\
\specialrule{.1em}{.05em}{.05em}
\multicolumn{2}{V{3} c V{3}}{\raisebox{13ex}{\makecell[c]{Schematics \\ of \\ physical process}}} &
			\includegraphics[width=33mm]{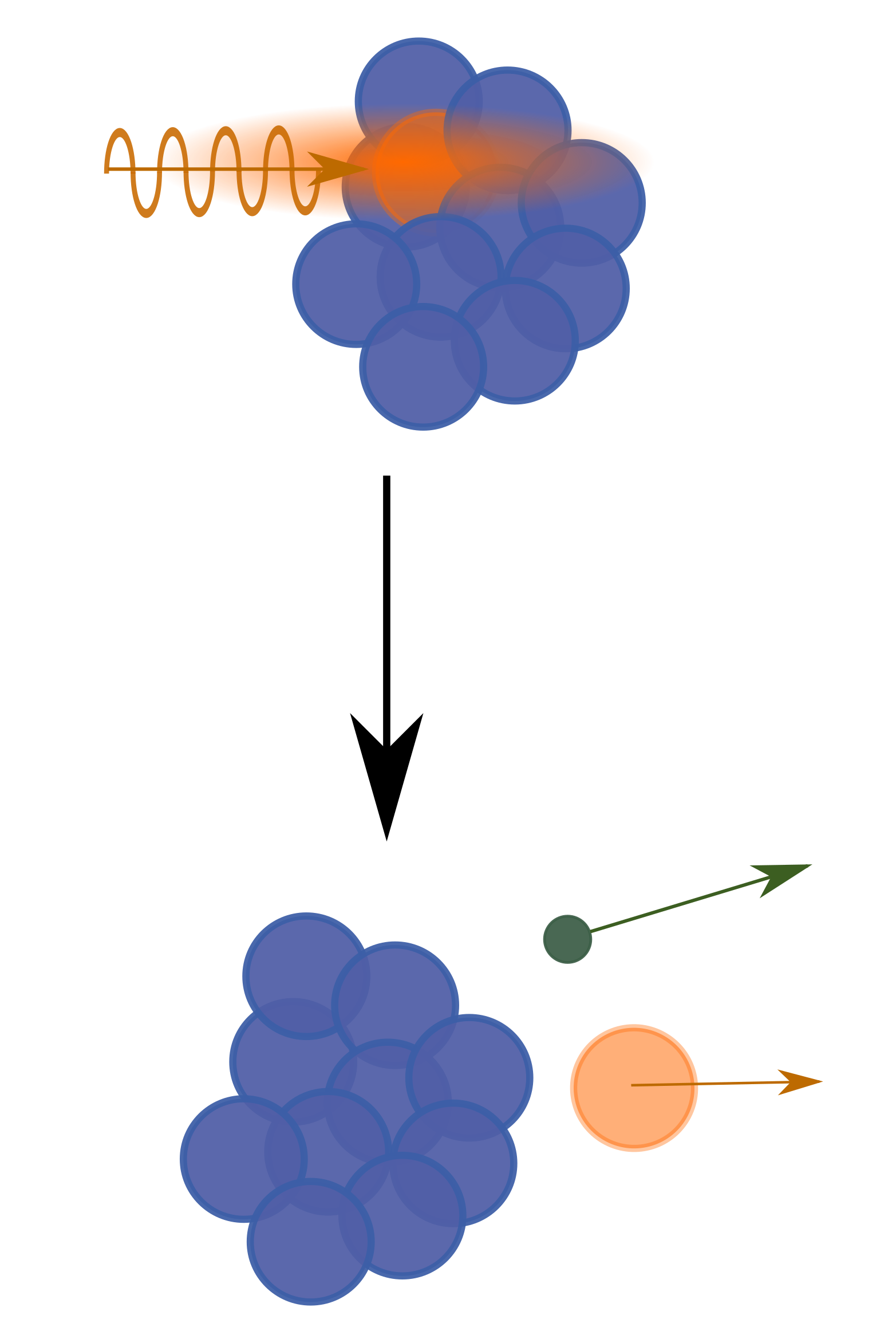} & 
			\includegraphics[width=33mm]{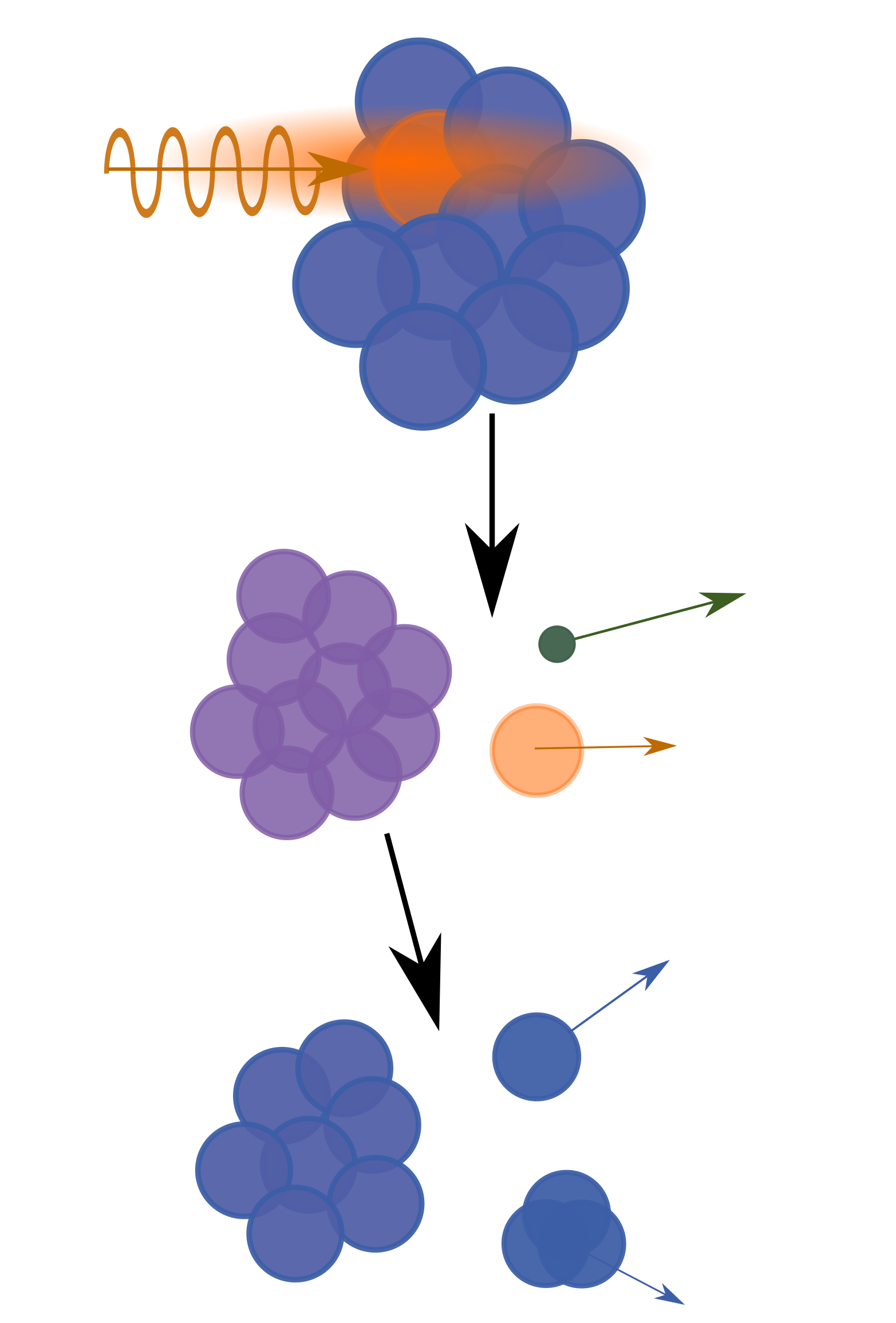} & 
			\includegraphics[width=33mm]{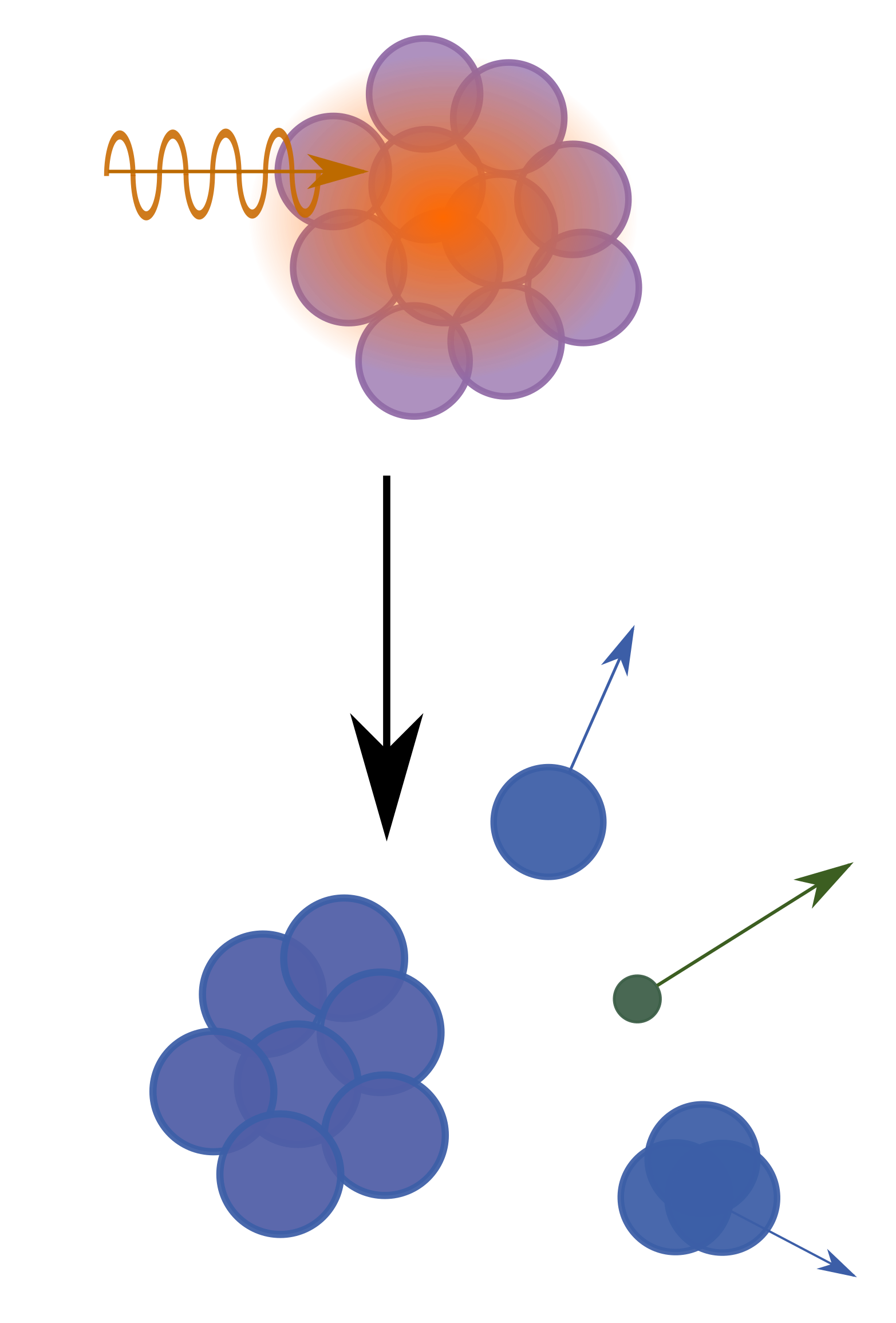} \\
\specialrule{.1em}{.05em}{.05em}
\multirowcell{2}{$\sigma_{A\gamma}$} & 
	\gape{\makecell[c]{$\rm \epsilon_r < 1\, GeV$}} & 
	\multirowcell{2}{$A \, \sigma^{\rm \textsc{Sophia}}_{p\gamma}$} & \multicolumn{2}{c V{3}}{$A^{\alpha}\,\sigma_{\rm univ}$, $\alpha=1$} \\\cline{2-2}\cline{4-5}
   &\gape{\makecell[c]{$\rm \epsilon_r > 1\, GeV$}} & & \multicolumn{2}{c V{3}}{$A^{\alpha}\,\sigma_{\rm univ}$, $\alpha=\alpha(\epsilon_r)$}\\
\specialrule{.1em}{.05em}{.05em}
\multirowcell{3}{$\sigma^{\rm incl}_{A\gamma \to X}$} 
	& \makecell[c]{Fragments \\ produced} & \makecell[c]{n/p and $A-1$} 
		& \makecell[c]{n/p and \\ disintegration \\ products at $E^*$ } & 
		\makecell[c]{$A_{\rm fr} \in [1-4,$\\$A/2 ... A-1]$\\(see Appendix~\ref{sub:reinjection_channels})}\\\cline{2-5}
	& \makecell[c]{Fragment`s \\ multiplicity}
		& \makecell[c]{Always 1 (for any \\ fragment produced)}
		& \makecell[c]{Combination of \\ disintegration model \\ and {\sc Sophia}}  & \makecell[c]{Empirical formulas\\and nuclear \\thermostatistics \\(see Appendix~\ref{sub:reinjection_channels})} \\\cline{2-5}
	& \makecell[c]{$\sigma^{\rm incl}_{A\gamma \to \pi}$} & \multicolumn{2}{c V{3}}{\gape{$A\,\sigma^{\rm incl,\, \textsc{Sophia}}_{p\gamma \to \pi}$}} 
	& \makecell[c]{$A^{\alpha_{\pi}}\,\sigma^{\rm incl}_{\pi}$\\(see Appendix~\ref{sub:photoproduction_of_pions_on_nuclei})}\\

\specialrule{.1em}{.05em}{.05em}
\end{tabular}
	\caption{Comparison of photomeson models and schematics of the related physical picture. 
	The nuclei and nuclear fragments are represented as collections of circles (nucleons) with 
	color related their role in the interaction (and fraction of the photon energy they receive): 
	blue, spectator nucleons  (no extra energy received, boost conservation); purple, non-active 
	participants (some extra energy received); and orange, active participants (receive most of the 
	energy, direct interaction). Green smaller circles represent pions.}
	\label{tab:model_table}
\end{table}

\subsection{Total photonuclear cross section} 
\label{sub:total_photonuclear_cross_section}

\begin{figure}
	\centering
	\includegraphics[width=.95\textwidth]{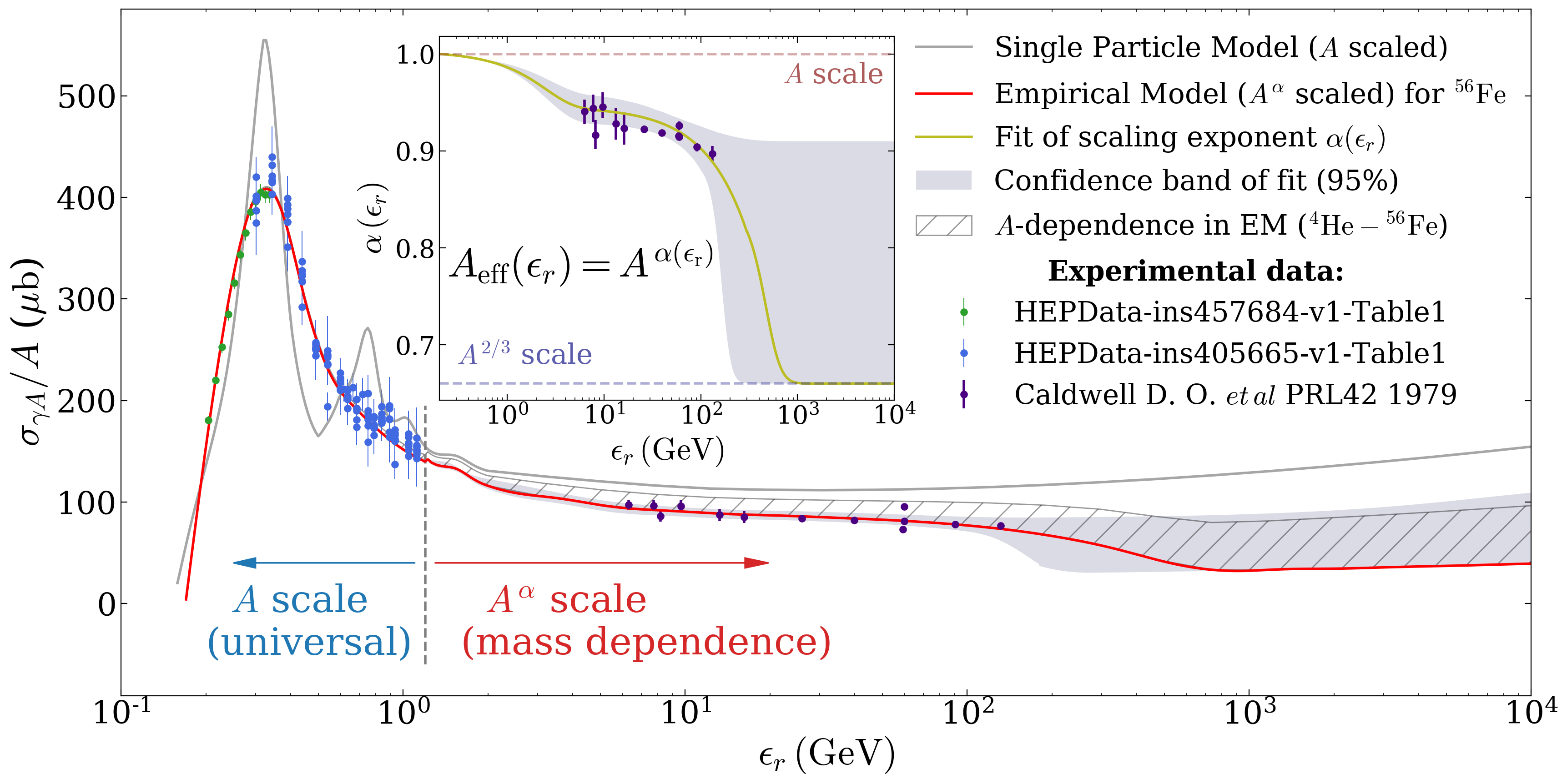}
	\caption{The inelastic photonuclear cross section divided by $A$ as a function of photon 
	energy in the nucleus' rest frame. Data points from a broad range of nuclei with 
	masses $A=7-208$ show a universal behavior at low energies and a mass dependence at higher
	energies. The red curve is the $\rm ^{56}Fe$ cross section as implemented in the EM, and 
	illustrates the typical shape in the model. The gray curve correspond to the SPM, common 
	to all nuclei and identical to the photo-nucleon cross section
	$\sigma^{\rm SPM}_{A\gamma}/A=\sigma_{N\gamma}$.
	The EM cross section has a mass scaling with $A^{\alpha}$ energy for $\epsilon_r > 1\, {\rm GeV}$ 
	(see \equ{cstot_EM_relations}). The dependence $\alpha(\epsilon_r)$ (yellow curve in the insert 
	plot) is a linear function fitted to data (confidence band shown as gray area). The extrapolation
	to lower energies is a sigmoid which takes the low energy value $\alpha_{\rm low}=1$.
	The extrapolation to higher energies is a sigmoid which tends to the theoretical limit
	$\alpha_{\rm lim}=2/3$).
	The hashed region represents the variation of the cross section per nucleon in EM 
	($\sigma^{\rm EM}_{A\gamma}/A=A^{\alpha - 1}\sigma_{N\gamma}$) for the mass 
	for the mass range $A=4-56$.}
	\label{fig:universal_function}
\end{figure}

As discussed in \sect{photon_interactions}, the nuclear environment (also referred to as 
``medium effects'') changes the physics above the pion threshold energies compared to the free 
nucleon case.
\figu{universal_function} shows the inelastic cross section divided by $A$ as a function of energy.
The curves corresponding to protons (SPM curve) and to $\rm ^{56}Fe$ (EM curve) are compared to 
photonuclear data compiled from experiments with various target
nuclei~\cite{MacCormick1997,Bianchi1996,Caldwell:1978yb}. 
At $\epsilon_r \lesssim \rm 1\,GeV$, the green circles correspond to light nuclei ($A=2-4$) and 
the blue circles to $A=7-208$. Within the errors of the data, a scaling with $A$ appears justified. 
However, the shape of the curve is different compared to $\sigma_{p\gamma}$, with only one 
pronounced resonance peak in place of the $P_{33}(1232)$ resonance ($\Delta$-resonance), being 20\% 
wider at half height and 30\% lower at the peak. The widening has been explained with the
$\rm Delta$-hole model~\cite{KOCH198499}, where medium effects are taken into account by including the 
Fermi motion of nucleons, the Pauli blocking restricting decay channels and 
the $\Delta$-N interactions. The nucleon resonances at energies beyond $500\,{\rm MeV}$ are not visible 
even for small masses such as Be and C~\cite{BIANCHI19935,PhysRevC.47.R922}.

For $\epsilon_r \gtrsim \rm 1\,GeV$ the photonuclear cross section per nucleon is reduced due to the 
shadowing effect. This reduction is dependent on the nuclear mass, and has been successfully understood 
within the Vector Meson Dominance model (VMD), which describes the photon's wave function as a superposition 
of mesonic states ($\rho$, $\omega$, $\varphi$) \cite{WEISE197453,PhysRevC.60.064617,Schildknecht:2005xr} that interact
hadronically with the nucleus. At higher energies, where the photon can resolve the partons in the 
nucleus, the parton distribution function is predicted to be high enough that the nucleus becomes
opaque to photons, leading to a theoretical limit similar to that of hadron-nucleus interactions
$\alpha_{\rm lim}=2/3$~\cite{Kaidalov:2001st}.
The mass scaling is typically parametrized in the literature as a power of the nuclear mass:
\begin{equation}
	\label{eq:cstot_EM_relations}
	\sigma_{A\gamma}(\epsilon_r)  = A_{\rm eff}(\epsilon_r) \, \sigma_{p\gamma}(\epsilon_r) = A^{\alpha(\epsilon_r)} \, \sigma_{p\gamma}(\epsilon_r).
\end{equation}
In the EM the cross section is parametrized parametrization with the following additional elements:
\begin{itemize}
	\item For $\epsilon_r < 1$ GeV a ``universal curve'' (spline fit of data points in 
	\figu{cross_sec_tot}) scaled by A ($\alpha=1$) is used. This reflects the universal shape of the cross section per nucleon exhibited by nuclei of a wide range of masses.
	\item For $\epsilon_r \ge 1$ GeV the photo-nucleon cross section is calculated with
	{\sc Sophia}, scaled by an energy dependent exponent $A_{\rm eff} = A^{\alpha(\epsilon_r)}$. 
	The energy dependence of the exponent is shown in the insert of \figu{cross_sec_tot}.
\end{itemize}

The data provided in \Ref~\cite{Caldwell:1978yb} are provided as 
$f_A(\epsilon_r)=A_{\rm eff}(\epsilon_r)/A$ for different nuclei (C, Cu, and Pb). Considering the 
parametrization with mass in \equ{cstot_EM_relations}, the exponent values $\alpha$ are calculated
from the data using $\alpha(\epsilon_r) = 1 + \ln{f_A(\epsilon_r)}/\ln{A}$ (purple points in the 
insert of \figu{universal_function}). A linear fit was performed to find the energy dependence, and
extrapolated towards the shadowing limit $\alpha_{\rm lim}=2/3$ for high energies (yellow curve in
insert).
The shaded area contains the 95\% confidence interval of the fit in the region where data are
available, and has also been extrapolated to higher energies: the upper band with a constant value,
the lower band with a transition towards $\alpha_{\rm lim}=2/3$ (\figu{universal_function}).

The mass scaling dependence with energy introduced in th EM at higher energies results in an $A$-
dependence of the cross section per nucleon $\sigma_{A\gamma}/A$. The hashed region in 
\figu{universal_function} represents this mass dependence for the range of masses $A=4-56$, where 
the upper line corresponds to the smaller mass and the lower to the larger mass.
	
\subsection{Photoproduction of pions off nuclei} 
	
	 \label{sub:photoproduction_of_pions_on_nuclei}

	 \begin{figure}[t]
	 	\centering
	 	\includegraphics[width=\textwidth]{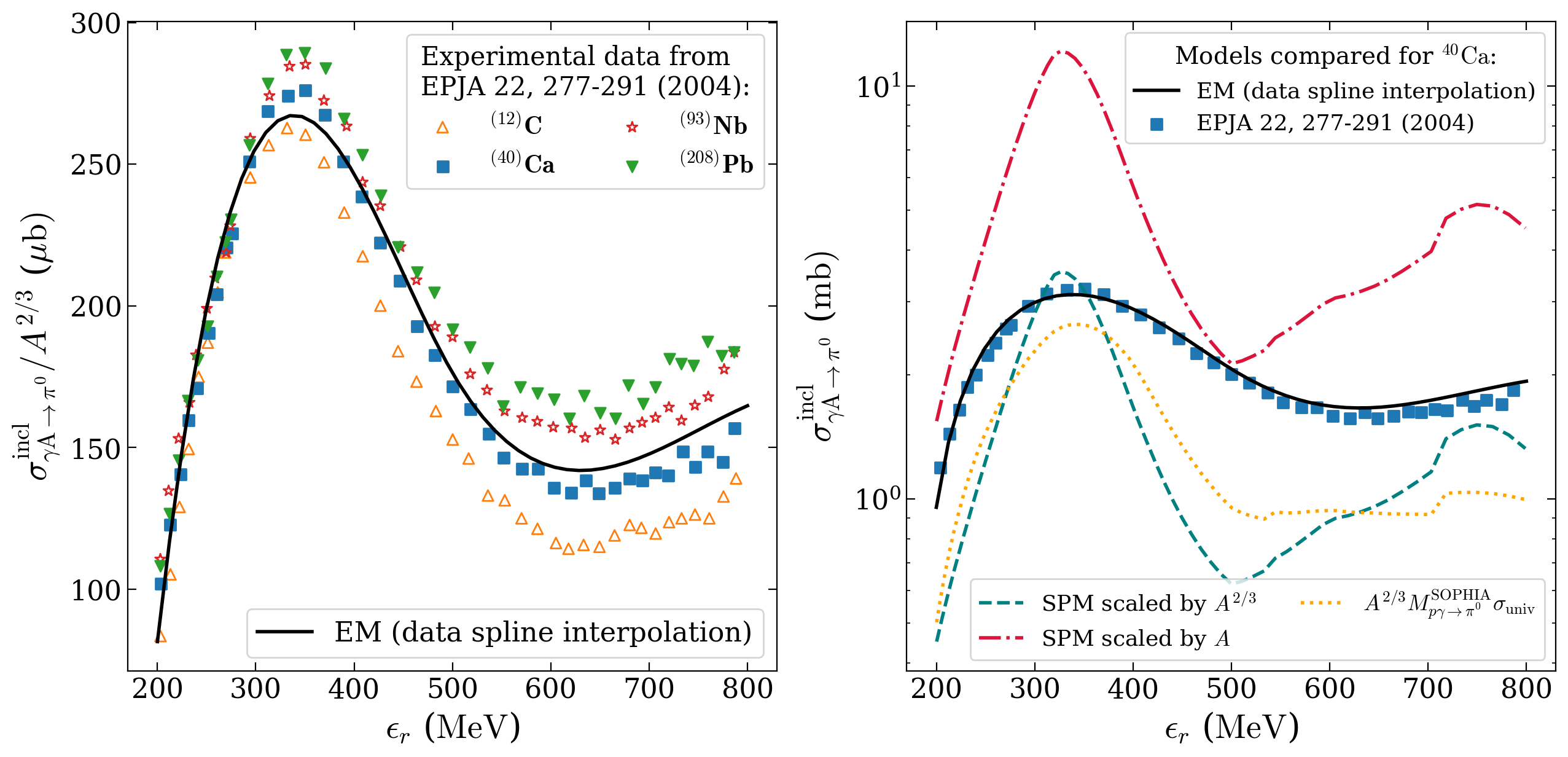}

	 	\caption{The figure shows the inclusive $\pi^0$ photoproduction cross section data 
	 	compared to different parametrizations. On the left, the cross sections are divided
	 	by $A^{2/3}$ as a function of the photon energy, and the data points are taken from Fig. 9 in
	 	\Ref~\cite{Krusche2004a}. The overall shape is similar for nuclei with different masses,
	 	and it is represented with a unique curve (solid black, spline interpolation of the data) which
	 	is implemented in the EM. On the right, the data and EM inclusive pion photoproduction off
	 	$\rm ^{40}Ca$ are shown with other parametrizations. The SPM curve with $A^{2/3}$ scaling
	 	(dashed blue) is too low, and with $A$ scaling (dot-dashed red) leads to overproduction.
	 	The total cross section per nucleon (universal curve $\sigma_{\rm univ}$) scaled by the 
	 	pion multiplicity (dotted orange) $M^{\rm \textsc{Sophia}}_{p\gamma \to \pi^0}$ is also not
	 	suitable to describe the data.}
	 	\label{fig:photopion_scaling}
	 \end{figure}

	The photoproduction of pions off nuclei is very sensitive to nuclear medium effects near threshold
	energies~\cite{Nagl:1991:NPP:2588247}. The interaction of the photon can occur via a ``quasi-free''
	process, where one nucleon and the pions produced are ejected while the rest of the nucleus is 
	unaffected. However, the pion(s) produced inside the nucleus interact with the surrounding nucleons 
	with a strong dependence on their kinetic energy. Pions interact weakly with nucleons inside the nucleus 
	for kinetic energies below $\sim 40 \, {\rm MeV}$, and exhibit a resonance around kinetic energies $50-100 \, {\rm MeV}$ 
	which depends on the nuclear mass~\cite{Lee&Redwine}. This effect is also evidenced in the
	photoproduction of pion pairs~\cite{Krusche2011}, where the mass scaling of the 
	cross section decreases from $A$ to $A^{2/3}$ as the photon energy becomes large enough to 
	produce pions with kinetic energies above $\sim 40 \, {\rm MeV}$. An effective description for 
	the mass scaling~\cite{Krusche2004} can be achieved by separating ``surface'' and ``volume'' 
	contributions, but such a description is hard to generalize for a broad range of nuclear masses.
	
	Pion photoproduction off nuclei can be calculated with existing codes using Monte Carlo techniques
	in a cascade scenario (CRISP~\cite{Deppman2004}), or in terms of transport equations 
	(GiBUU~\cite{Buss2012}). These simulations can be set up to have a good overall agreement with 
	data. However, a general table for a wide range of nuclear species is computationally expensive 
	and at the same time, the details of pion emission from each species 
	are usually less important in astrophysical simulations compared to the uncertainties of nuclear composition. 
	To keep the EM description of the pion photoproduction off nuclei as simple as in 
	the SPM, the same shape for the inclusive cross section is used for all nuclei with a 
	mass dependence and scaling coefficient $\alpha_{\pi}(\epsilon_r)$.

	\figu{photopion_scaling} (left) shows the inclusive $\pi^0$ photoproduction as a function of 
	incident photon energy $\epsilon_r$ for energies below $\sim 1\, {\rm GeV}$. The data were 
	measured for different nuclei and normalized to $A^{2/3}$ (values reproduced from Fig.~9 in
	\Ref~\cite{Krusche2004a}). The similarities in the curves point to a dominating ``surface''-like
	mass scaling ($A^{2/3}$) rather than a ``volume''-like mass scaling ($A$). The systematic 
	differences between the different curves are due to second order processes which scale with volume 
	like multi-pions produced at low kinetic energies~\cite{Krusche2004}. The data have been
	fit with a spline (black solid curve), which represents the inclusive pion production cross section
	fit to all nuclear masses assuming a scaling with $A^{2/3}$.
	\figu{photopion_scaling} (right) compares the $\rm ^{40}Ca$ pion production cross
	section with different parametrizations. The SPM curve with $A$ scaling (dot-dashed red line) predicts on
	average more than twice the number of pions compared to data (blue squares), whereas when using 
	$A^{2/3}$ scaling (dashed blue) it leads to less than half. The total cross section per nucleon 
	(universal curve $\sigma_{\rm univ}$) scaled by the pion multiplicity per nucleon 
	$M^{\rm \textsc{Sophia}}_{p\gamma \to \pi^0}$ and a $A^{2/3}$ mass scaling (dotted orange) underestimates
	data, as well. The spline fit to all available data in \figu{photopion_scaling} (left) scaled by 
	$A^{2/3}$ describes the pion production sufficiently well.
	
	At higher energies heavier mesons start playing a role in the final states, affecting the simplified scaling 
	assumption. For example, studies of	photoproduction of $\eta$ mesons up to $2 \, {\rm GeV}$~\cite{Mertens2008} find 
	consistency with a $A^{2/3}$ scaling when the $\eta$ is produced with kinetic energies sufficiently above threshold, 
	demonstrating strong absorption through FSI. These type of effects are currently studied but the theoretical 
	description is not yet complete~\cite{Dieterle,Kaeser2016}.	However, as the photon energy increases such 
	effects are less important, and have been shown to become irrelevant for photon energies beyond
	$50 \; {\rm GeV}$~\cite{PhysRevC.60.064617} where the production of pions is consistent with $A$ scaling.

	Our inclusive description with a mass scaling coefficient does not capture the complexity of the exclusive
	final states. Hence, our model follows the available data that suggests a $A^{2/3}$ scaling near threshold
	with a common curve for all species of nuclei, and is extended to higher energies
	by allowing the exponent to increase towards the limit value 1.

	The pion photoproduction off nuclei in the EM is, therefore, parametrized as 
	$A^{\alpha_{\pi(\epsilon_r)}}\sigma^{\rm incl}_{\pi}$ where $\sigma^{\rm incl}_{\pi}(\epsilon_r)$ is an 
	inclusive cross section for $\pi^+$, $\pi^-$ and $\pi^0$ production, as described below.
	Below $1 \; {\rm GeV}$ $\sigma^{\rm incl}_{\pi}(\epsilon_r)$ is derived from data: the form of $\sigma^{\rm incl}_{\pi^0}$ 
	is obtained by fitting a spline to the experimental values (black curve in \figu{photopion_scaling}). 
	For charged pions, their multiplicities are adjusted proportional to the production ratio off nucleons 
	as taken from {\sc Sophia}:
	\begin{equation}
	\label{eq:pion_relation}
	\sigma^{\rm incl}_{\pi^{\pm}}(\epsilon_r) = 
		\frac{M^{\rm \textsc{Sophia}}_{N\gamma \to \pi^{\pm}}(\epsilon_r)}
		{M^{\rm \textsc{Sophia}}_{N\gamma \to \pi^0}(\epsilon_r)}
		\sigma^{\rm incl}_{\pi^0}(\epsilon_r) \; .	
	\end{equation}
	
\begin{figure}[t]
		\centering
		\includegraphics[width=.8\textwidth]{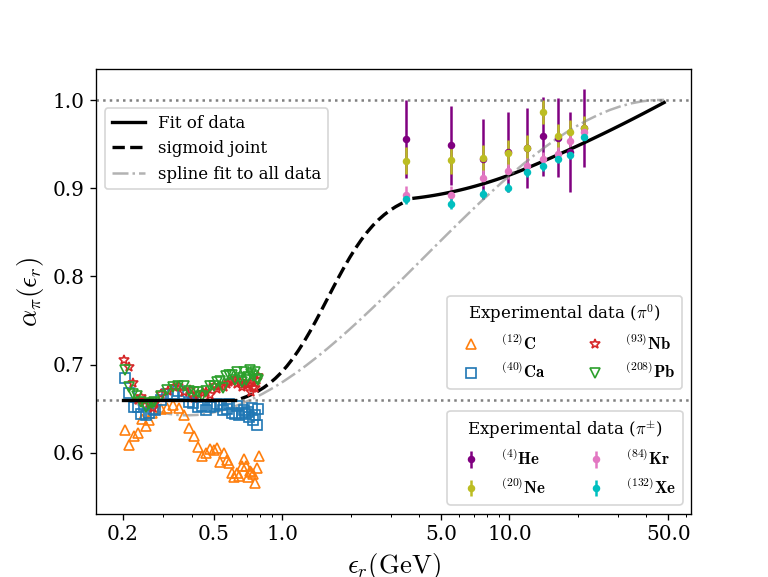}
		
		\caption{Energy dependence of the scaling exponent in the EM parametrization of the 
		pion production inclusive cross section. The data values have been calculated from
		hadron multiplicity ratios reported in \cite{Airapetianetal.2001,AIRAPETIAN200337,AIRAPETIAN20071}.
		The production of different pion species is even~\cite{Airapetianetal.2001,AIRAPETIAN200337}, therefore
		we use data from all of them to derive a unique $\alpha_{\pi}(\epsilon_r)$.}
		\label{fig:pion_alpha}
	\end{figure}

	Above $1 \; {\rm GeV}$ the EM curve is extended as in the SPM but with a smooth transition 
	from $A^{2/3}$ scaling to $A$ scaling with photon energy (see \figu{pion_alpha}) at around $50 \; {\rm GeV}$:
	\begin{equation}
		\sigma^{\rm incl}_{A\gamma \to \pi} =A^{\alpha_{\pi}(\epsilon_r)}\sigma^{\rm incl}_{\pi} = 
		\begin{cases}
			A^{\alpha_{\pi}(\epsilon_r)}\sigma^{\rm \textsc{Sophia}}_{N\gamma \to \pi} &
				{\rm for}\enspace 1\,{\rm GeV} < \epsilon_r < 50\,{\rm GeV} \\
			A \sigma^{\rm \textsc{Sophia}}_{N\gamma \to \pi} &
				{\rm for}\enspace \epsilon_r > 50\,{\rm GeV}
		\end{cases}	\; ,
		\label{eq:cs_pi_HE}
	\end{equation}
	satisfying also the relations in \equ{pion_relation}.

	As demonstrated in \figu{pion_alpha} the energy dependence of the pion scaling exponent $\alpha_{\pi}(\epsilon_r)$
	is derived from experimental data~\cite{Airapetianetal.2001,AIRAPETIAN200337,AIRAPETIAN20071}.
	The data has been obtained in deep inelastic scattering experiments measuring the semi-inclusive hadron production
	from the interaction of virtual photons with nuclei. The reported magnitude is the multiplicity ratio, which 
	describes ratio of pion production per nucleon in a nucleus $X$ to that in Deuterium $D$ (see 
	\equ{multiplicity_ratio_definition}).
	\begin{equation}
		R^{\pi}_M (\epsilon_r) = \left( \frac{\sigma^{\rm incl}_{A_X \to \pi}}{A_X} \right) / 
								\left(  \frac{\sigma^{\rm incl}_{A_D \to \pi}}{A_D} \right) .
		\label{eq:multiplicity_ratio_definition}
	\end{equation}
	It follows from \equ{cs_pi_HE} that the EM the multiplicity ratio is related to the photon energy 
	nucleus mass and the scaling exponent as:
	\begin{equation}
		R^{\pi}_M (\epsilon_r) = \left( \frac{A_X}{A_D} \right)^{\alpha_{\pi}(\epsilon_r) - 1} .
		\label{eq:multiplicity_ratio_A_dependence}
	\end{equation}
	Below $1 \; {\rm GeV}$ the scaling exponent 
	is fixed to $2/3$ as suggested by data in \figu{photopion_scaling}. The multiplicity ratio data, from different nuclear species, show a sharp increase of the scaling 
	exponent in the interval from $1-3 \, {\rm GeV}$. A spline fit to the entire energy range
	(dash-dotted gray curve in \figu{pion_alpha}) underestimates data from $4-10 \, {\rm GeV}$. For this 
	reason the model joins the the two energy ranges below 3 and $4-22 \, {\rm GeV}$ using a 
	sigmoid function (dashed black line). This curve applies to all pion species, since
	they are produced in equal amounts~\cite{AIRAPETIAN200337}.

	The energy redistribution of pions (differential cross sections) (see \equ{redist_function})
	used are those sampled from {\sc Sophia} for free nucleons, \ie{} any modification to the angular 
	distributions resulting from the nuclear medium is neglected.

	\subsection{Nuclear fragmentation} 
		\label{sub:reinjection_channels}
		
		In the SPM a photonuclear interaction of a nucleus with mass $A$ always results in the loss of one 
		nucleon per interaction. Hence, the final state always includes a remnant nucleus of mass $A-1$ that 
		does not further disintegrate. In the photomeson regime nuclear breakup is governed by several 
		mechanisms that lead to a more complex final state, including the emission of multiple nucleons, 
		light fragments and a distribution of remnant nucleus masses. We consider two physical approaches to 
		model this effect:

		\textbf{Residual Decay Model (RDM)}: The initial photon-nucleon interaction is modeled in 
		the same way as in the SPM, assuming a quasi-free target nucleon. The motion of the nucleon through 
		the nuclear medium results in an average energy loss, proportional to the nuclear radius 
		(Abrasion-Ablation hypothesis).
		This energy left in the residual nucleus with mass $A-1$ is an excitation energy which can
		been estimated as $\epsilon^* = 17 \, {\rm MeV} (A-1)^{1/3}$ \cite{Rachen:1996ph}. The 
		subsequent de-excitation proceeds through a re-scattering, or an equivalent interaction a 
		photon with the energy $\epsilon_r = \epsilon^*$ with the remnant nucleus ($A-1$). The 
		typical energies for the range of masses are close to the GDR regime, for which the 
		multiplicity distributions for final state particles have been obtained from a statistical, 
		Hauser-Feshbach based model {\sc Talys} \cite{talys18} that uses combinations different 
		nuclear models to describe the disintegration. These tabulated cross sections have been 
		previously discussed in \Ref~\cite{boncioli2016nuclear}. The inclusive cross section in the RDM 
		for producing the fragment $k$ from the interacting species $j$ is calculated by 
		normalizing the multiplicity of the fragments produced in the disintegration model at
		$\epsilon^*$ to the total cross section of the interacting species $\sigma_j$ 
		\begin{equation}
		\label{eq:RDM_cs_inclu}
		\sigma^{\rm incl}_{j \to k}(\epsilon_r) = \sigma_j(\epsilon_r) \left( \delta_{\rm kN} +
			\frac{\sigma^{\rm \textsc{talys}}_{(j-1) \to k}(\epsilon^*)}
			{\sigma^{\rm \textsc{talys}}_{(j-1)}(\epsilon^*)} \right),
		\end{equation}
		where $\delta_{\rm kN}$ (1 if k is proton or neutron and zero otherwise) accounts for the 
		additional nucleon emitted in the first photon interaction.

		\textbf{Empirical Model (EM)}: The photonuclear interaction is more complex, and often 
		involves multiple nucleons and intermediate fragments. The production of nuclear 
		fragments is very dependent on the absorption mechanism and the consequent energy transfer
		to the rest of the nucleus (intra-nuclear cascade). The average production of masses can
		be modeled using empirical formulas previously obtained in \Ref~\cite{1402-4896-49-3-004}
		and reproduced in Appendix~\ref{sub:formulas_for_the_cross_sections}. The formulas estimate 
		the average cross sections for producing certain fragments in the energy range
		$0.2 \lesssim \epsilon_r \lesssim 1\, {\rm GeV}$. Employing those relations and additional considerations from thermostatistics, the inclusive cross sections for a broad range 
		of nuclear fragments have been produced (see Appendix~\ref{sub:formulas_for_the_cross_sections}). 
		Schemes of the physical scenarios and a summary of the components of these models are 
		shown in the \Tab~\ref{tab:sources_parameters} in comparison with the SPM.

		\begin{figure}[t]
			\includegraphics[width=7.5cm]{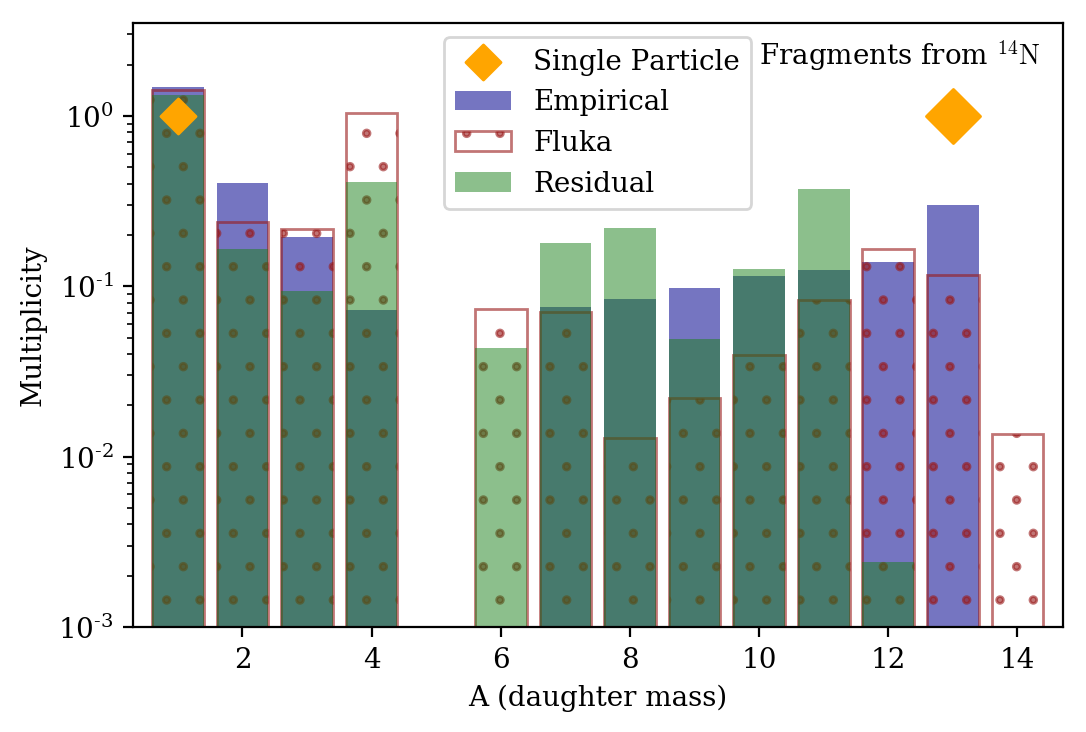}
			\includegraphics[width=7.5cm]{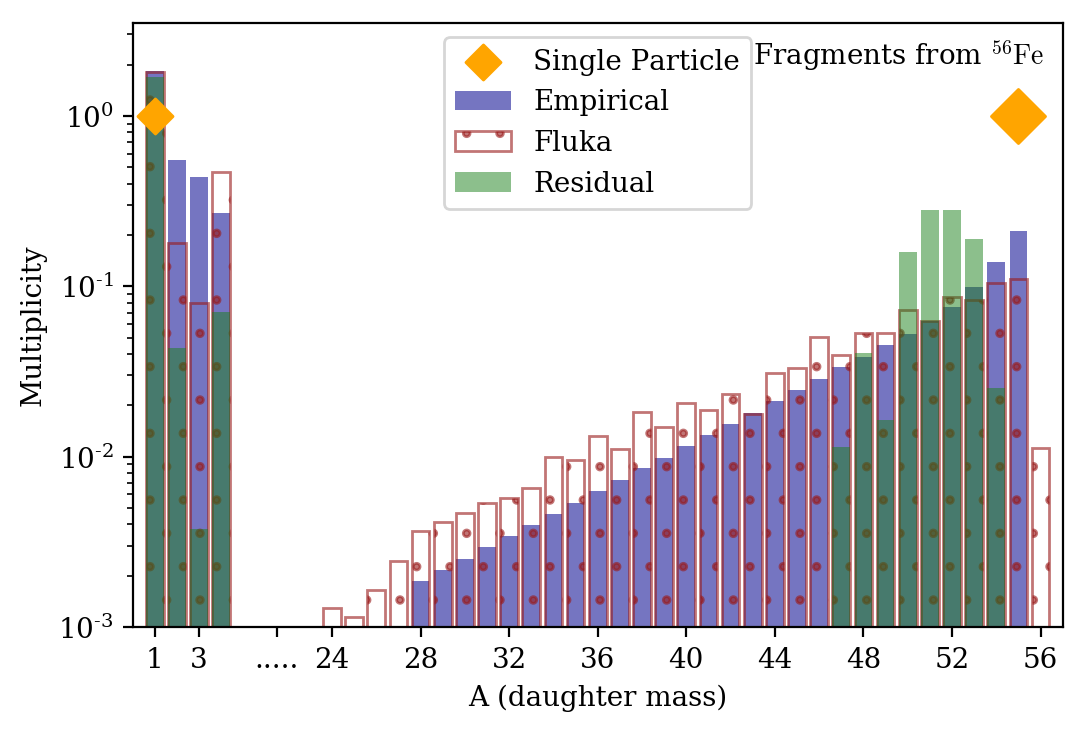}

			\caption{The mass distribution of the daughter fragments for different interacting nuclei
			$\rm ^{14}N$ (left) and $\rm ^{56}Fe$ (right).
			The multiplicity is the average number of particles produced of certain species 
			(see \equ{multiplicity}). The distribution corresponding to the 
			Residual Decay Model (green) is peaked as is characteristic of evaporation-dominated 
			disintegration. The distribution corresponding to the Empirical Model (blue) is more 
			spread its shape is in better agreement with that obtained in {\sc Fluka}~\cite{fluka} which 
			is a detailed a Monte Carlo code based on data and theory (red outline). 
			In the SPM masses $1$ and $A-1$ (orange diamonds) are always produced from species with 
			mass $A$.}
			\label{fig:model_multiplicities}
		\end{figure}

		\begin{figure}[t]
			\centering
			\includegraphics[width=.8\textwidth]{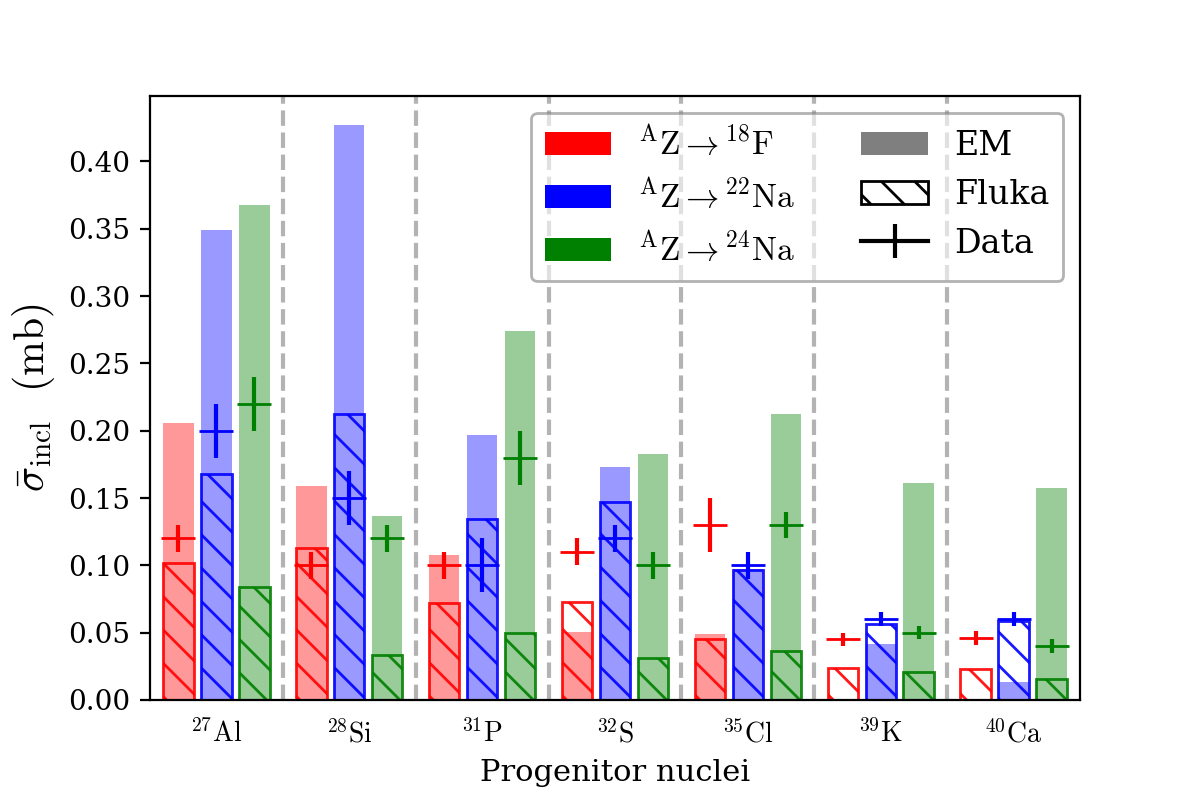}

			\caption{Comparison of the cross sections averaged over the range $\rm 0.3-1\, GeV$
			obtained from {\sc Fluka}, the EM and experimental data from \cite{DINAPOLI19743633}.
			EM performs similar to {\sc Fluka}. Note that both models are have not
			been specifically optimized for these particular isotopes.}
			\label{fig:EM_fragments_vs_data}
		\end{figure}

		In \figu{model_multiplicities} a comparison between of the mass distribution 
		of fragments is shown for two interacting nuclei: $\rm ^{14}N$ (left) and $\rm ^{56}Fe$ (right).
		The multiplicities are calculated from the total cross section of the interacting
		species $j$ and the inclusive cross section for production of species $k$:
		\begin{equation}
			M_{j \to k} = \frac{\sigma^{\rm incl}_{j \to k}}{\sigma_j}
		\end{equation}
		which is equivalent to \equ{multiplicity} and represents the mean number of particles of a 
		certain species produced per interaction. The species are grouped by mass and the 
		multiplicity of the mass group is the sum of the multiplicities of all species with the 
		same mass. The EM values are shown in solid blue, and the RDM in solid green. 
		Additionally, the values obtained with {\sc Fluka}~\cite{fluka} are shown as reference.
		In {\sc Fluka}, detailed Monte Carlo calculations are combined with state of the art theory and
		nuclear data to	simulate the photonuclear interaction and the subsequent production of 
		secondaries. The shape of the EM is in better overall agreement with {\sc Fluka} than the RDM.
		The implicit assumptions in the RDM are reflected in the narrower mass distribution 
		which is a typical shape produced in the disintegration at lower energies around the GDR,
		where the disintegration model for the residual nucleus is sampled. Experimental results 
		for the disintegration at energies corresponding to the photomeson regime point to
		processes like spallation and fission~\cite{DINAPOLI10.106,DiNapoli1978} (see
		\figu{partial_processes}) which are explicitly simulated in {\sc Fluka} whereas in 
		EM the mass distribution is fixed for all species of the same mass. \figu{EM_fragments_vs_data} 
		shows a comparison of EM and {\sc Fluka} to data from \cite{DINAPOLI19743633} where the average production over photon energies $\rm 0.3-1\, GeV$
		is measured for multiple nuclei. The corresponding values from the EM and {\sc Fluka} are within a factor 1-3 of the 
		experimental values. For this case, the EM performs similarly to a more elaborate nuclear code.
		The SPM can not be even compared in the same way, since it produces only one $A-1$ fragment.

		The differential cross sections (energy redistributions) of the fragments from the statistical
		fragmentation in the RDM and EM follow the relation in \equ{delta_func_per_nucleus} (boost conservation).
		The redistribution for secondaries produced directly in the photon-nucleon interaction are sampled
		from {\sc Sophia} (see \equ{redist_function}). A fully detailed description of the model's implementation
		is located in Appendix~\ref{sec:photomeson_model_based_on_empirical_formulas}.



\section{Impact in astrophysical scenarios} 
	\label{sec:source_scenarios}

	The characteristics of our new model impacts the production of UHECRs and neutrinos, depending on
	the parameters of the sources. We show in the following two
	representative example sources where the importance of the photomeson production have been already
	highlighted: GRBs~\cite{Biehl2018a} and TDEs~\cite{Biehl2018b}. The examples have been chosen such
	that photomeson processes dominate the disintegration at the highest energies.

	GRBs are the most energetic electromagnetic outburst class. Shells of plasma emitted by the
	central engine can create internal shocks that become acceleration sites for charged particles. The
	interactions of those particles with the target photons in the so-called prompt phase
	can result in the production of a significant number of neutrinos. As a consequence, GRBs
	are candidate sources for UHECRs and neutrinos, although existing analyses already constrain
	some regions of the parameter space for GRBs~\cite{Aartsen:2017wea} multi-messenger models.
	
	In this work we use an example from \Ref~\cite{Biehl2018a}, where the interactions in the
	source are simulated with the NeuCosmA code for nuclei in the context of a one-zone model,
	meaning that the collistions of plasma shells cluster at the same collision radius $R$.
	Provided that baryons are present in the source, the density of the radiation triggers
	a nuclear cascade in which nuclei lighter than the primary are produced due to
	photodisintegration and photomeson production, impacting the mass composition of in-source and
	ejected cosmic rays. In \Ref~\cite{Biehl2018a}, several qualitative cases have been distinguished:
	for a high radiation density in the source, the nuclear cascade strongly develops (``optically thick case'')
	and produces a high flux of nucleons, which dominate the neutrino production. If the radiation
	density is low (``empty cascade''), the nuclear cascade does not develop and the neutrino flux
	is dominated by photomeson production of the primary nuclei. In the intermediate case (``populated cascade''),
	the nuclear cascade develops and neutrinos are efficiently produced off primary and secondary
	nuclei. The different cases can be quantitatively distinguished by the optical thickness to
	photohadronic interactions at the highest energies.

	These different cases relating the degree disintegration to the multi-messenger production 
	as introduced in \Ref~\cite{Biehl2018a} for GRBs, can be applied to other source classes~\cite{Biehl2018b}.
	One example are jetted Tidal Disruption Events (TDEs), which have been proposed as possible
	UHECR sources \cite{Farrar:2014yla}. In this scenario, a star is gravitationally disrupted in the
	vicinity of a black hole by tidal forces, generating a jet in which a nuclear cascade similar the
	GRB can develop. 
	
	The source parameters used in the considered models are listed in
	\Tab~\ref{tab:sources_parameters}. The luminosity $L$ and collision radius $R$ define the
	photon energy density in the source, together with the Lorentz factor of the shells $\Gamma$. The
	duration $T$ can vary a lot among different source classes, with consequences for the detection
	capability of different experiments. The baryonic loading $\xi$ is defined as the energy injected
	as baryons over the energy injected in photons, and it is fixed to a reference value here. 
	The efficiency of the acceleration controls the maximum energy of acceleration. The spectrum of the
	GRB is typically described by a broken power law with a spectral break at 1 keV in the observer's
	frame and spectral indices $\alpha = -2/3$ $(-1)$ and $\beta = -2$ below and above the break energy
	for TDEs (GRBs), respectively. Although the photodisintegration through excitation of the GDR is
	the dominant process for UHECRs with respect to photomeson processes, the interplay of the spectral
	index of the photons with the GDR and photomeson regimes may favor the photomeson production at the
	highest energies, as for example in TDEs. The minimal cutoff of the photon
	spectrum in the source can also drastically reduce the GDR interaction length, as demonstrated
	already in \cite{Biehl2018a}. The critical parameters influencing photomeson production at the
	highest energies are marked boldface in the table.

	\begin{table}[t]
		\centering
		\begin{tabular}{|l|c|c|c|}
			\hline
			Source & TDE \cite{Biehl2018b} & GRB \cite{Biehl2018a} \\[.3em] 
			\hline \hline
			Gamma factor & $\rm \Gamma =  10$ & $\rm \Gamma = 300$ \\[.3em] 
			Redshift & $z = 0.001$ & $z = 2$ \\[.3em] 
			Duration & $T = 10^6 \, \rm s $ & $T = 10 \,\rm s$ \\[.3em] 
			Luminosity & $ L_X = 10^{47}\, \rm erg/s$ & $ L_{\gamma} = 10^{53}\, \rm erg/s$ \\[.3em] 
			Collision Radius & $ R = 10^{9.6}\, \rm km$ & $ R = 10^{8.3}\, \rm km$ \\[.3em] 
			Injected isotope & $\rm ^{14}N$ & $\rm ^{56}Fe$ \\[.3em] 
			Acceleration efficiency & $\eta = 1$ & $\eta = 1$ \\[.3em] 
			Baryonic loading & $\xi = 10$ & $\xi = 10$ \\[.3em] \hline
			\multirow{5}{4.5cm}{Target photon spectrum parameters in the shock rest frame} & 
			  $\varepsilon'_{X,{\rm br}}=1\,{\rm keV}$  & $\varepsilon'_{\gamma,{\rm br}}=1\,{\rm keV}$ \\[.3em] 
			   & $\varepsilon'_{X,{\rm min}}=10^{-6} \,{\rm eV}$ &
			     $\boldsymbol{\varepsilon'_{\gamma,{\rm min}}=100\,{\rm eV}}$\\[.3em] 
			   & $\varepsilon'_{X,{\rm max}}= 300 \,{\rm keV}$ &
			     $\varepsilon'_{\gamma,{\rm max}}= 300 \,{\rm keV}$ \\[.3em] 
			   & $\boldsymbol{\alpha = - 2/3}$, $\beta=-2$ & $\alpha = - 1$, $\beta=-2$\\[.3em] 
			\hline
		\end{tabular}
		\caption{The source parameters shown were taken from the references 
		\citep{Biehl2018a, Biehl2018b}. In the case of the GRB, the parameters for 
		the Optically Thick scenario are shown; the ones for the other scenarios 
		(Empty Cascade and Populated Cascade) are identical except for their luminosities 
		$L_{\gamma} = 10^{49} \rm erg/s$ and $L_{\gamma} = 10^{51} \rm erg/s$, respectively. Parameters critical for the photomeson processes dominating at the highest energies are highlighted in boldface.}
		\label{tab:sources_parameters}
	\end{table}
	
\begin{figure}[t]
			\centering
			\includegraphics[width=\textwidth]{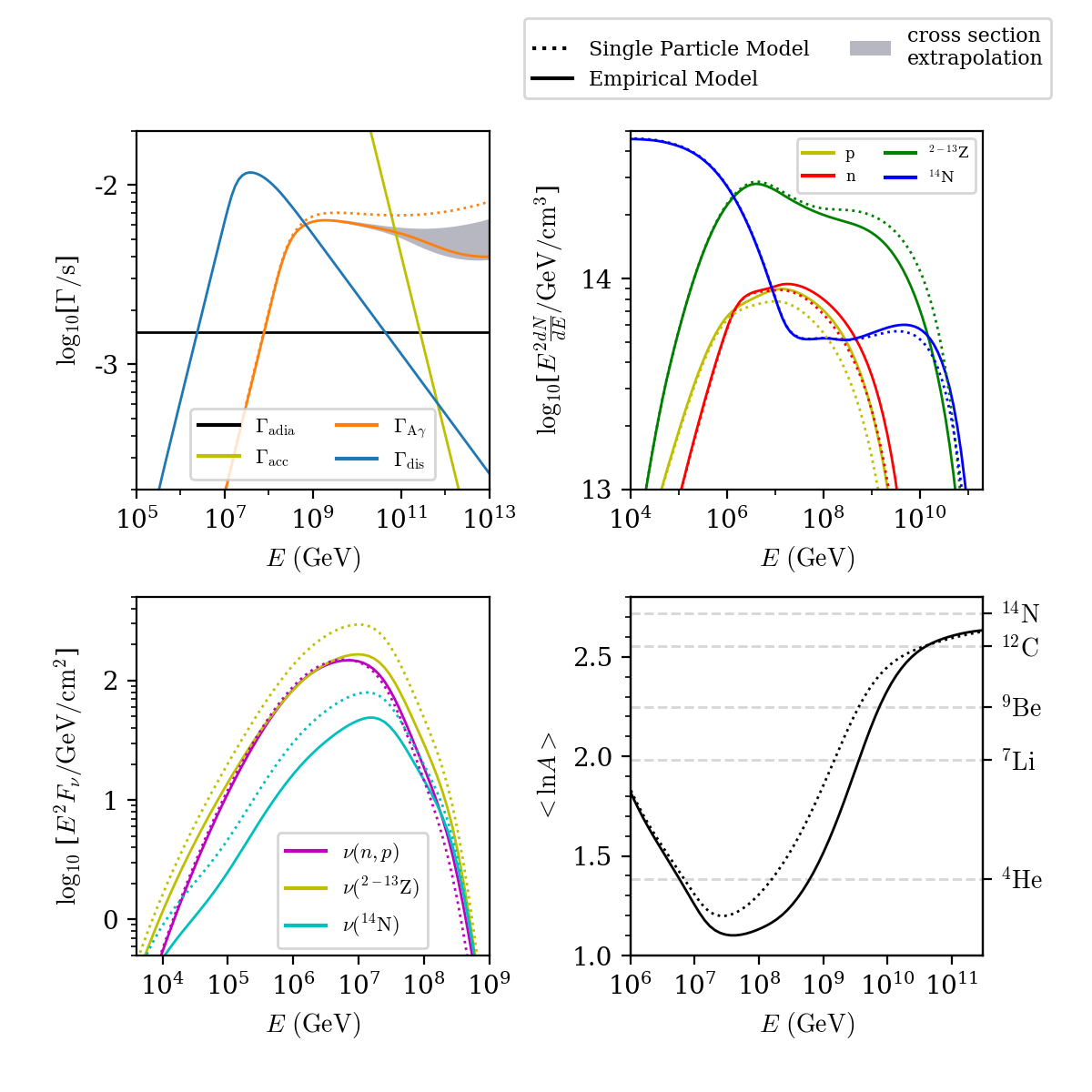}
			\caption{Comparison of the results using the new Empirical Model (EM) (solid curves) 
			with the reference Single Particle Model (SPM) (dotted curves) in the TDE scenario from
			\Ref~\citep{Biehl2018b}; shaded areas refer to uncertainty in the cross section 
			extrapolation. The different panels show: the leading process rates for the injected
			$\rm ^{14}N$ (upper left panel), the in source densities (upper right panel), the flux
			of neutrinos grouped by origin (lower left panel), and the averaged $\ln A$ of the 
			in-source spectra, all as a function of the energy $E$ in the observer's frame.}
			\label{fig:TDE_impact}
		\end{figure}

	Let us focus on the TDE scenario first, see \figu{TDE_impact}. The slope of the photon spectrum at
	low energies reduces the photodisintegration rate at high energies (high energy protons interact
	with low energy photons), which means that the photomeson regime dominates the photohadronic
	interactions at the highest energies -- as it is shown in the upper left panel of \figu{TDE_impact}.
	Here the impact of the cross section systematics is only significant beyond the maximal energy
	of the cosmic rays, which can be estimated by balancing the acceleration rate with the sum of 
	the energy losses (acceleration is possible only when losses are subdominant). In the
	upper right panel, the effect of the lower cross section adopted in the EM with respect to the SPM
	is visible: the injected primary nuclei (blue) is less depleted in the EM, therefore less energy
	is injected in secondaries. On the other hand the the production of secondaries is distributed
	over a larger range of masses (see \figu{model_multiplicities}) and the average mass of the 
	secondaries is smaller than in the SPM, which implies a more efficient disintegration. The 
	increase in the nucleons densities is the result of this more efficient disintegration of all 
	the secondaries produced.

	The neutrino production, see lower left panel, is mostly affected by the cross section scaling of
	the pion production; the heavier the primary, the larger the difference between the models. It is
	noteworthy that the SPM changes the qualitative observation that the neutrino production would be
	dominated by interactions of nuclei: In the EM, both contributions are similar in magnitude but nuclei
	dominate slighlty at the highest energies. Overall the neutrino flux is depleted by a factor of $~1.5$ if the EM is
	used with respect to the SPM. 
	The effect of the disintegration channels is appreciable in the $< \ln A >$ where the EM leads to
	a smaller cascade mass (lower right panel). This happens mainly at the intermediate energies 
	range because of the conservation of the Lorentz factor of the nuclei in the 
	photodisintegration process, and because the secondaries are produced at lower energies in 
	proportion to their mass ratio to progenitor species (due to boost conservation). At the highest 
	energies, where the primary nuclei density dominates, the composition is slightly heavier than in SPM due to the reduced interaction.         

	\begin{figure}[t]
			\centering
			\includegraphics[width=\textwidth]{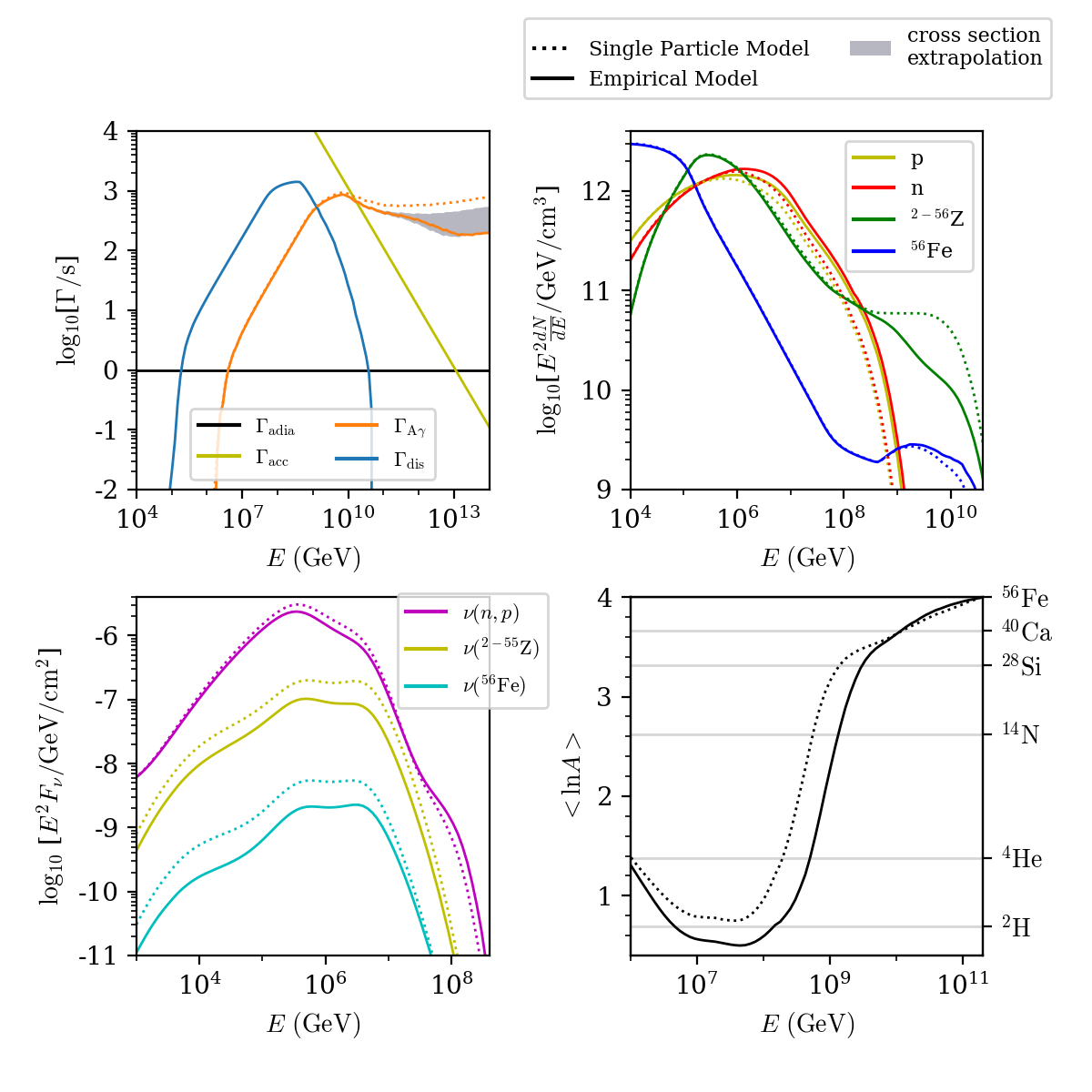}
			\caption{Similar to \figu{TDE_impact}, but for the GRB scenario with parameters shown 
			in \Tab~\ref{tab:sources_parameters} from \Ref~\citep{Biehl2018a}, for the injected
			$\rm ^{56}Fe$.} 
			\label{fig:GRB_impact}
		\end{figure}

		\begin{figure}[t]
			\centering
			\includegraphics[width=1.0\textwidth]{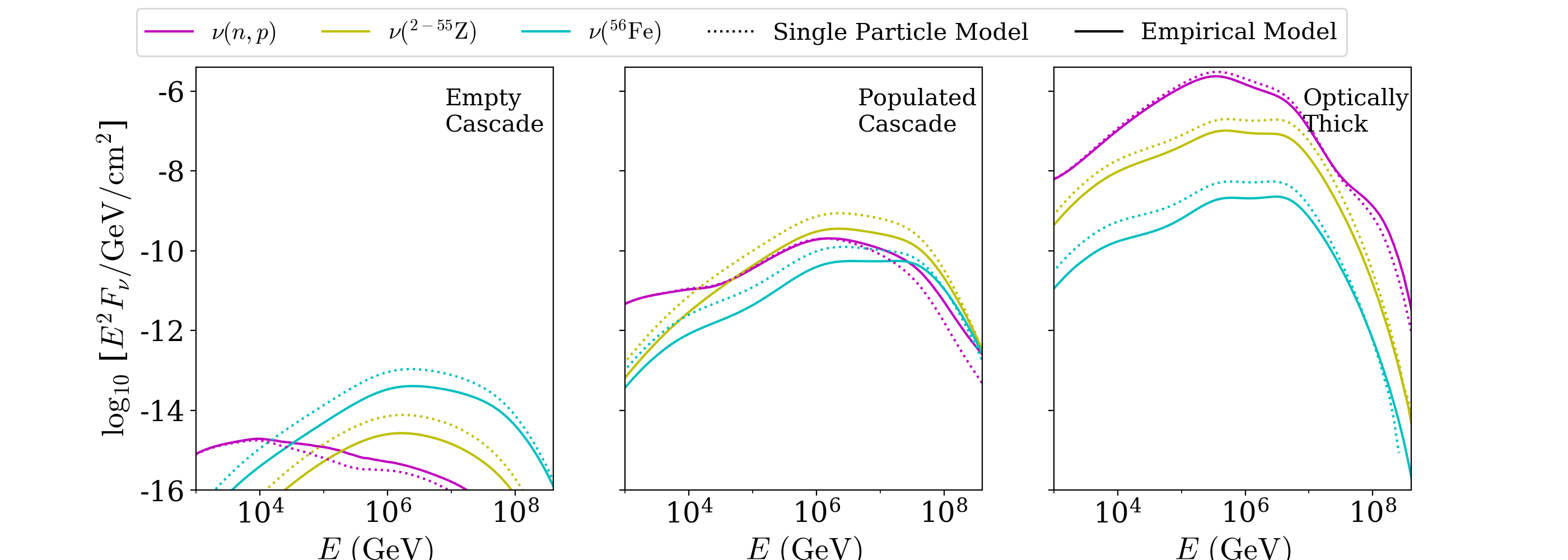}
			\caption{Neutrino production corresponding to different GRB luminosities and to SPM 
			(dotted curves) and EM models (solid curves) for photomeson process.}
			\label{fig:GRB_luminosity_comparison}
		\end{figure}
		
	For the GRB example, we follow \Ref~\citep{Biehl2018a}. The luminosity and the collision radius reported
in \Tab~\ref{tab:sources_parameters} correspond to a high radiation density, and the neutrinos are
mainly given by the secondary nucleons produced in the nuclear cascade. In particular, we use the
assumptions of Appendix~B in \Ref~\citep{Biehl2018a} for the photon spectrum of the source, in which
a higher low-energy cutoff of the photon spectrum is investigated resulting in an suppression of
photodisintegration rate with respect to the photomeson rate. The upper left panel of \figu{GRB_impact}
demonstrates how the changes in the cross section affect the interaction rates at energies above
the maximum energy of acceleration, which means that the effects on the energy densities (upper right panel)
are dominated by the redistribution channels of the secondaries. The impact on the more efficient nuclear
disintegration (through a larger number of open channels) is also
visible in the $\langle \ln A \rangle$ behavior (bottom right) which overall exhibits a lighter composition, 
reaching up to three times lower value compared to the SPM. The neutrino production (bottom left) is again 
dominated by the pion production scaling used in EM  ($\rm A^{2/3}$) in contrast with SPM scaling 
($\rm A$), and the effect is large because of the heavier primary mass compared to the TDE example.

\figu{GRB_luminosity_comparison} shows the impact of the change of the photomeson model in
the neutrino fluxes, corresponding to different GRB luminosities identified with different nuclear
cascade regimes.  Note that, compared to the main text of \Ref~\cite{Biehl2018a}, here the low energy
photon cutoff has been applied to all examples.  The largest differences are visible in the low and
intermediate luminosities, for which the neutrinos are mainly produced off (primary and secondary)
nuclei. In the populated-cascade scenario (center), the smaller neutrino yields from the
intermediate nuclei in the EM, change the leading production channel of the neutrinos. When using SPM
in this intermediate case, the neutrino emission from secondary nuclei dominates the flux, whereas
for the EM the contribution from nucleons is comparable to that from secondary nuclei.

\section{Summary and conclusions} 
\label{sec:conclusions}

	We have studied photomeson production off nuclei in astrophysical environments, where photomeson
	interactions refer to photon-nucleus collisions at photon energies $\gtrsim 140 $ MeV in the 
	nucleus' rest frame. Above this threshold, the production of secondary mesons commences, and hence
	the production of astrophysical neutrinos in the multi-messenger context. Our main focus has been
	the treatment of the nuclear interactions in astrophysical environments for which the interacting
	nuclei have large kinetic energies resulting in forward peaked the secondary spectra. We have
	scrutinized a commonly adopted approach in the literature known as Single Particle Model (SPM), and
	introduced an improved description: the Empirical Model (EM). The impact of the EM was studied
	in astrophysical scenarios in which the photomeson processes are known to dominate at the highest
	energies (certain classes of GRBs and TDEs).
		
	The SPM treats the nucleus as superposition of nucleons scaling the nucleus the total inelastic
	cross section with $A$ or $A^{2/3}$. We have discussed the disagreements of this model with 
	available data, and improve those in the EM by using: a data-driven parametrization for the total
	inelastic cross section at low energies and a mass number scaling consistent with data and theory
	for high energies; a data-driven parametrization of the pion photoproduction cross section resulting
	in a $A^{2/3}$ scaling at the lower energies that accounts for nuclear medium effects and final state 
	interactions; a nuclear breakup model for the remnant nucleus based on existing empirical formulae for 
	partial photo-emission mechanisms, resulting in a better agreement with distributions obtained from
	Monte-Carlo simulations, such as {\sc Fluka}.
	
	These modifications affect the photomeson interactions off nuclei in astrophysical environments,
	and we used two examples from the literature in which the photomeson regime is known to dominate
	the nuclear interactions at the highest energies. Our GRB example resembles an effect from
	synchrotron self-absorption in which low energy target photons are suppressed that would
	otherwise disintegrate the UHECRs via Giant Dipole Resonance excitation. Our TDE example has
	a broken power law target photon spectrum with a hard enough (low) power law index such that the
	photomeson production dominates at the highest energies. 
	
	For these cases, we demonstrated that the improved model affects the nuclear cascade in the source
	resulting in an ejected UHECR composition that is up to three times lighter, and a reduction of the
	neutrino flux by up to 50\%.  The nuclear cascade is affected by the cross section (such as in the
	TDE case) or the additional channels in fragmentation (such as for our GRB case). The impact on the
	neutrino flux stronger (as in SPM) depends on the mass of the isotope(s) dominating the neutrino
	production, implying that neutrino emission becomes more sensitive to the choice of the injection
	composition into an internal-shock GRB model and the degree of the nuclear cascade. For high radiation
	densities in the optically thick case, neutrino production is dominated by nucleon interactions, and
	hence the impact from the new model is low. Consequently, the strongest effects will occur for
	the populated and empty nuclear cascade cases with heavy injection isotopes, which, however, have
	a smaller neutrino production efficiency compared to cases in which the
	cascade fully develops.
	
	The impact on
	the UHECRs is smaller and only expected in environments where the photomeson regime dominates the
	nuclear disintegration. A prominent counter-example is, for instance, cosmic ray transport
	in the extragalactic space. In a rigidity-dependent model for the spectra of different nuclear
	species emitted from the accelerating source, it is found that the UHECR spectrum and composition
	are best-fitted by a low rigidity cutoff \cite{Aab:2016zth,AlvesBatista:2018zui,Heinze:2019jou},
	meaning that the photomeson production is not triggered and the nuclear breakup occurs
	predominantly via photodisintegration off the extragalactic photon backgrounds. The corresponding
	cosmogenic neutrino flux at the main peak is dominated by interactions of the UHECRs with the
	cosmic infrared background (because of the low rigidity), and the contribution of nucleon and
	nuclei interactions in \Ref~\cite{Heinze:2019jou} at the peak is about 50-50. This means that the
	neutrino flux may be even lower (by up to about 25\%) for our photomeson model.
	
	We finally  note that our model is based on a very small number of parameters and can be easily
	implemented for cross comparisons in any of the existing codes and calculation frameworks for
	radiation modeling of High Energy Particle Astrophysics. The tools for reproducing the model are 
	available~\cite{leonel_morejon_2019_2600177} and can be used for implementation in any framework
	simulating the interaction of UHECRs.

\acknowledgments

	We would like to thank J. Rachen, J. Heinze and A. van Vliet for useful discussions.

	This project has received funding from the European Union's 
	Horizon 2020 research and innovation programme under grant 
	agreement no. 646623.

\begin{appendix}

\section{Photomeson model based on empirical formulas} 
	\label{sec:photomeson_model_based_on_empirical_formulas}

	To construct the photomeson model it is necessary to obtain for each nuclear species $j$:

	\begin{itemize}
		\item $\sigma_{j} = \sigma_{j}(\epsilon_r)$ the absorption photonuclear 
		cross section as function of the photon energy in the nucleus rest frame $\epsilon_r$.
		\item $\sigma^{\rm incl}_{j \to k} = \sigma^{\rm incl}_{j \to k}(\epsilon_r)$ the inclusive 
		photonuclear cross section for producing each of the $\rm k$-th possible product particles
		(pions, nucleons, nuclear fragments). 
	\end{itemize}

	The form of $\rm \sigma_{j}$ in the EM model is presented in
	\Sec~\ref{sub:total_photonuclear_cross_section} and summarized in \Tab~\ref{tab:model_table}. 
	The expressions for the inclusive cross sections are presented in
	\Sec~\ref{sub:photoproduction_of_pions_on_nuclei} for pions and in
	\Sec~\ref{sub:reinjection_channels} for nucleons and nuclear fragments.
	The following sections detail the calculation of $\sigma^{\rm incl}_{j \to k}$ for nuclear 
	fragments. The energy redistribution of the secondaries is calculated using \equ{delta_func_per_nucleus} 
	for the fragments of the nuclear remnant, while for direct interactions of the photon
	with a nucleon it is obtained from {\sc Sophia} using \equ{redist_function}
	(see Appendix~\ref{sub:small_fragment_yields_from_statistical_mechanics}).

	\subsection{Formulas for the inclusive cross sections} 
	\label{sub:formulas_for_the_cross_sections}
		
		The inclusive cross sections for producing nucleons and nuclear fragments are based on the
		formulas discussed in sections below. The following relations are used in our model:
		\begin{align}
			\sigma^{\rm incl}_{j \to {\rm p}} &= \sigma^{\rm dir}_{\rm p} + \sigma^{\rm sp}_{\rm p} \; ,\\
			\sigma^{\rm incl}_{j \to {\rm n}} &= \sigma^{\rm dir}_{\rm n} + \sigma^{\rm sp}_{\rm n} + 
                 \sum_{x = 2}^{x_{\rm max}}x\sigma^{\rm mul}_{x{\rm n}}\;,
                 	\enspace x_{\rm max} = \lfloor 1.4 \, A^{0.457} \rfloor \label{eq:cs_n_incl} \; ,\\
			\sigma^{\rm incl}_{j \to k} &= 
				\begin{cases}
					\sigma^{\rm dir}_{\rm p} & {\rm for\; nucleus}\; k=(Z-1, A-1)\\
					\sigma^{\rm dir}_{\rm n} & {\rm for\; nucleus}\; k=(Z, A-1)\\
					\sigma^{\rm mul}_{x{\rm n}} & {\rm for\; nucleus}\; k \in (Z, A-x)\\
					\sigma^{\rm sp}_k & {\rm for\; nucleus}\; k\in (Z_k\in[1,...,Z-1], A_k\in[1,...,A-2])
				\end{cases} \; ,
				\label{eq:cs_k_incl}
		\end{align}
		\noindent where the expressions for $\sigma^{\rm dir}_{\rm p}$, $\sigma^{\rm dir}_{x {\rm n}}$,
		$\sigma^{\rm mul}_{\rm n}$, and $\sigma^{\rm sp}_k$ are obtained using empirical relations, 
		 derived from data in the energy range $\rm 0.2-1\, GeV$ in \Ref~\cite{1402-4896-49-3-004}.
		 Here we adapt the naming convetion and remark that these relations do not
		 correspond to the microscopic nuclear processes, but instead give a reasonable representation
		 of data.


		\begin{figure}[t]
			\centering
			\includegraphics[width=0.75\textwidth]{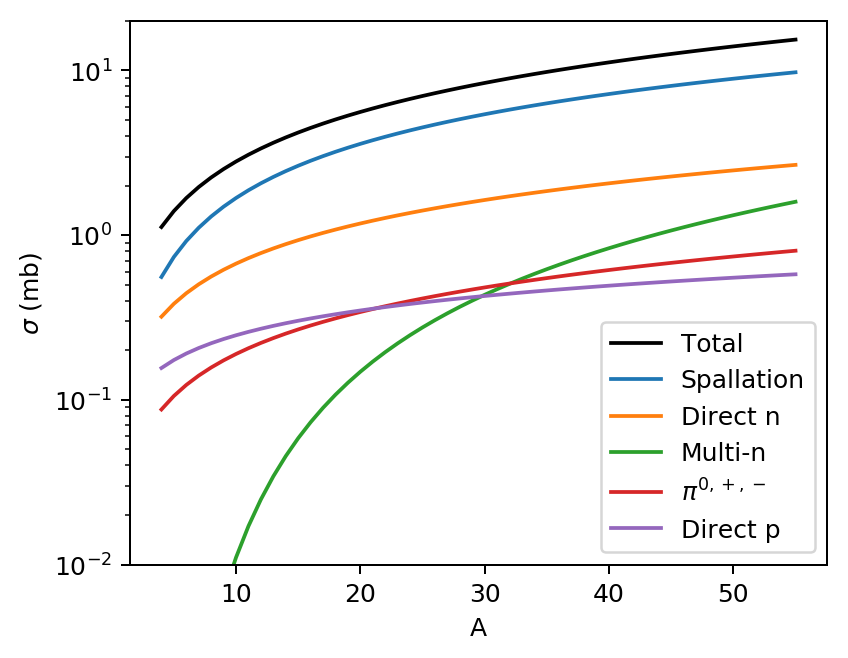}
			\caption{Cross sections (in milibarn) versus the mass of the interacting nucleus $A$, 
			for the relevant processes included in the EM. This is a partial reproduction of 
			Fig. 8 in \Ref~\cite{1402-4896-49-3-004}.}
			\label{fig:partial_processes}
		\end{figure}

		For simplicity and due to the sparsity of data, we assume they hold constant for all photon
		energies in the photomeson regime (see \figu{cross_sec_tot}).
		Below are the processes and their expressions reproduced from \Ref~\cite{1402-4896-49-3-004}:

		\begin{itemize}
			\item \textbf{Direct proton production ($\sigma^{\rm dir}_{\rm p}$)} Reactions 
			producing only one proton $(\gamma, \rm p)$ and a residual nucleus of mass $A-1$:
			\begin{equation}
				\sigma^{\rm dir}_{\rm p} = 0.078 \, A^{0.50} \; {\rm mb} \; .
				\label{eq:csp}
			\end{equation}
			\item \textbf{Direct neutron production ($\sigma^{\rm dir}_{\rm n}$)} Reactions 
			producing only one proton $(\gamma, \rm n)$ and a residual nucleus of mass $A-1$:
			\begin{equation}
				\sigma^{\rm dir}_{\rm n} = 0.104 \, A^{0.81} \; {\rm mb}  \; .
				\label{eq:csn}				
			\end{equation}
			\item \textbf{Multi-neutron production ($\sigma^{\rm mul}_{x {\rm n}}$)} Reactions 
			producing $x>1$ neutrons $(\gamma, x \rm n)$ and the corresponding residual nucleus 
			with mass ${\rm A} - x \rm n$:
			\begin{align}
				\sigma^{\rm mul}_{\rm ntot} &= \sum_{x = 2}^{x_{\rm max}}
				\sigma^{\rm mul}_{x{\rm n}}\; {\rm mb} ,\;\,{\rm where}\; 
				x_{\rm max} = \lfloor 1.4 \, A^{0.457} \rfloor \; {\rm and} \\
				\sigma^{\rm mul}_{x{\rm n}} &
				= 0.187 \, A^{0.684} \, e^{-37 \, A^{-0.924}\,(x\, -\, 1)^{5/4}}\; {\rm mb} \; .
				\label{eq:csxn}
			\end{align}
			\item \textbf{Spallation ($\sigma^{\rm sp}_{x {\rm p} y {\rm n}}$)} Spallation 
				reactions where a nominal loss of $x>1$ protons and $y>1$ neutrons occurs
				$(\gamma, x {\rm p}, y {\rm n})$ and the corresponding residual nucleus with mass
				${A_r = A} - x {\rm p} - y {\rm n} \geq A/2$ is produced:
			\begin{align}
				\sigma^{\rm sp}_{{x {\rm p} y {\rm n}}} &= 
				15.7 \, \mathsf{E}^{-1.356} \, e^{-3.03 \, \mathsf{E}^{-1.06}(x-1) - 0.466 \, 
				(x\, -\, C \, \alpha \, y)^2} \; \rm mb  
				\label{eq:csSpall}\; , \;where \\
				\mathsf{E} &=446/{\rm A} \,,\; C\, =\, 2.3 \, \alpha - 1.044 \;, \; {\rm and}
				\; \alpha= Z/(A-Z) \nonumber \; .
			\end{align}
			\item \textbf{Pion production ($\sigma^{\rm prod}_{\pi}$)} Reactions 
			producing pions ($\pi^0, \pi+, \pi^-$) and one nucleon $(\gamma, \pi + \mathcal{N})$.
			This contribution is only used in the EM for normalizing the spallation in \equ{cs_spall_mean}.
			The relations for pion production in the EM are discussed in 
			\Sec~\ref{sub:photoproduction_of_pions_on_nuclei}:
			\begin{equation}
				\sigma^{\rm prod}_{\pi} = 0.027 \, A^{0.847} \; {\rm mb} \; .
				\label{eq:cspi}
			\end{equation}
			\item \textbf{Fission (not considered here)} For the nuclei considered in this work 
			(mass up to $\rm ^{56}Fe$) this process has no values or experimental 
			data~\cite{1402-4896-49-3-004} but needs to be included for much higher masses.
		\end{itemize}
		
		These relations are represented graphically in \figu{partial_processes}, except for the spallation curve which is 
		strongly dependent on the mass and nucleons emitted, thus $\sigma^{\rm sp}$ is represented as the
		remainder of the total cross section $\sigma^{\rm tot}=0.28 A \; {\rm mb}$ (see
		 \Ref~\cite{1402-4896-49-3-004}) after subtracting all contributions

		\begin{equation}
			\sigma^{\rm sp} = \sigma^{\rm tot} - \left(\sigma^{\rm prod}_{\pi} + 
			\sigma^{\rm dir}_{\rm p} + \sigma^{\rm dir}_{\rm n} + \sigma^{\rm mul}_{\rm ntot}\right) \; .
			\label{eq:cs_spall_mean}
		\end{equation}

		To obtain the inclusive cross sections used in EM it is necessary to multiply the formulas above by the 
		number of respective particles produced. For example, in the inclusive contributions from direct proton and 
		direct neutron production ($\sigma^{\rm dir}_{\rm p}$ and $\sigma^{\rm dir}_{\rm n}$) only one nucleon
		and one large fragment ($Z-1$, $A$) or ($Z$, $A-1$) are produced. In the case of the multi-neutron process
		the cross section $\sigma^{\rm mul}_{x {\rm n}}$ for the emission of $x$ neutrons contributes to the
		inclusive cross section for neutron emission as $x\sigma^{\rm mul}_{x {\rm n}}$ (see \equ{cs_n_incl}), \ie{}:
		\begin{equation}
			\sigma^{\rm mul}_{\rm n} = \sum_{x = 2}^{x_{\rm max}}x\sigma^{\rm mul}_{x{\rm n}} \; .
			\label{eq:neutron_inclusive}
		\end{equation}

		In the case of spallation, only one fragment species $k$ with mass $A/2 \le A_k \le A-2$ is produced and the 
		inclusive cross section is just $\sigma^{\rm sp}_{x {\rm p} y {\rm n}}$. We have made additional considerations
		to group the produced nucleons ($x$ protons and $y$ neutrons) into fragments with masses $A_k \in [1..4]$

		\begin{align}
			\sigma^{\rm sp}_k = 
				\begin{cases}
					\sigma^{\rm sp}_l \;& A_k \in [1...4]\\
					\sigma^{\rm sp}_{x {\rm p} y {\rm n}} \;& A_k \in [A/2... A-2]
				\end{cases} \; .
			\label{eq:cs_spall_k}
		\end{align}

		Where $\sigma^{\rm sp}_l$ stands for any light fragment produced 
		($\sigma^{\rm sp}_{\rm n}$, $\sigma^{\rm sp}_{\rm p}$, \etc). The method of estimating
		$\sigma^{\rm sp}_l$ is discussed in the following subsection.


	\subsection{Inclusive cross section of small fragments} 
		\label{sub:small_fragment_yields_from_statistical_mechanics}

		In the processes considered in \Sec~\ref{sub:formulas_for_the_cross_sections} no fragments with masses smaller 
		than $A/2$ are created. We assume for the EM that nucleons produced in the spallation process can be grouped 
		into fragments of no more than four nucleons and estimated the number of those fragments with 
		thermostatistical formulas~\cite{Cole2010}.

		For any spallation event $i$ a number of $x$ protons and $y$ neutrons is lost from the interacting nucleus 
		($Z$, $A$) with an exclusive cross section $\sigma_{x{\rm p}y{\rm n}}$ (which in the following will be labeled
		as $\sigma^{\rm sp}_i$). We consider that the total energy 
		of left from the interaction is taken by the spalled nucleons as their kinetic energy (no internal excitation 
		of products). The spalled nucleons can be configured in a number of small fragments, or a combination $r$:
		\begin{equation}
			{\rm C}^r_i = \left\{{\rm N}^r_{i,{\rm n}}, {\rm N}^r_{i,{\rm p}}, ..., {\rm N}^r_{i,l}, 
				..., {\rm N}^r_{i,{\rm \alpha}} \right\}
			\enspace l \in (Z_l, A_l), \enspace 1 \le A_l \le 4 \; ,
		\end{equation}
		where $l$ refers to any of the nuclear species with no more than four nucleons and with a decay half life 
		longer than the relevant timescale of the astrophysical (only common isotopes remain).

		The set of all possible $\mathbb C_i=\{{\rm C}^r_i\}$ is determined by finding all mixtures of species
		$l$ in the quantities ${\rm N}^r_{i,l}$ such that the proton and nucleon number of the combination matches 
		the spalled numbers:
		\begin{align}
			x = Z_i = \sum_{{\rm C}^r_i} Z_l {\rm N}^r_{i,l} \; , \label{eq:normalization_Z}\\
			x+y = A_i = \sum_{{\rm C}^r_i} A_l {\rm N}^r_{i,l} \; . \label{eq:normalization_A}
		\end{align}

		An appropriate weight for each combination ${\rm P}_{i,r}$ allows to find the numbers of particles 
		produced in each spallation event $i$ with
		\begin{equation}
			{\rm N}_{i,l} = \sum_r {\rm N}^r_{i,l}{\rm P}_{i,r} \; .
			\label{eq:number_of_l}
		\end{equation}
		By summing the numbers of particle $l$ over all spallation configurations, the total inclusive cross section
		of spallation for particle $l$ is obtained
		\begin{equation}
			\sigma^{\rm sp}_l = \sum_{i} {\rm N}_{i,l} \; \sigma^{\rm sp}_i  \; ,\enspace \enspace
			l \in {\mathbb S} \; .
			\label{eq:cs_spall_l}
		\end{equation}

		\noindent The following section details the estimation of the combination weights ${\rm P}_{i,r}$.


	\subsection{Evaluation of the weights of combinations} 
		\label{sub:evaluation_of_the_weights_of_combinations}
		
		The simplest assumption is that all combinations are equiprobable, that is they all have the same 
		weight ${\rm P}_{i,r} = {\rm P}_i$. Then the number of particles of species $l$ (\equ{number_of_l}) 
		results 
		\begin{equation}
			{\rm N}_{i,l} = {\rm P}_i\sum_r {\rm N}^r_{i,l} \; ,
			\label{eq:equiprob_numbers}
		\end{equation}
		and the ${\rm P}_i$ can be found from the nucleon conservation. This means that the number of nucleons 
		produced in all spallation events is equal to the nucleons present in the interacting nucleus
		\begin{equation}
			A\sum_i \sigma^{\rm sp}_i = \sum_k A_k \sigma^{\rm sp}_k \; ,
			\label{eq:nucleon_conservation}
		\end{equation}
		where $\sigma^{\rm sp}_k$ are those in \equ{cs_spall_k} and \equ{cs_spall_l} with \equ{equiprob_numbers}.

		The equiprobability assumption fails to account for the fact that combinations with more stable nuclei are
		more likely to occur. In the EM we apply statistical mechanics, assuming that the combinations are possible
		microstates corresponding to a certain spallation event (macrostates) within the Grand Canonical distribution. 
		A general form of the partition function in this case is~\cite{Cole2010}
		\begin{equation}
			\mathsf Z = \sum_{i, r} e^{-\beta(\epsilon_i - \mu n_r)} \; ,
		\end{equation}
		where $\epsilon_i$ is the energy of the macrostate, $\mu$ is the chemical potential of the elements 
		(here only one species is included) and $n_r$ the number of elements of a given microstate. The weight 
		of a microstate can be expressed by the probability of the microstate
		\begin{equation}
			{\rm P}_{i,n} = e^{-\beta(\epsilon_i - \mu n_r)} / {\mathsf Z} \; ,
		\end{equation}

		We are interested in the relative weights of different microstates with the same energy
		$\epsilon_i$ since we use the normalization condition \equ{nucleon_conservation}.
		Hence, the expression for ${\rm P}_{i, n}$ can be rewritten grouping the
		factors common to a macrostate $i$ into some normalization constant $\mathsf B_i$
		\begin{equation}
			{\rm P}_{i, n_r} = {\mathsf B_i}e^{\beta \mu n_r} \; ,
		\end{equation}
		which allows estimating the weights of the macrostates without calculating the partition function.
		Considering each nuclear species $l$ as a different constituent of the system, the chemical potential 
		associated will be its ground state energy $\mu_l = m_l$ (rest mass in units of energy) which can be 
		found in nuclear mass tables. Then the required expression of ${\rm P}_{i, r}$ is
		\begin{equation}
			{\rm P}_{i,r} = {\mathsf B_i}e^{\beta \sum\limits_{l} \mu_l {\rm N}^r_{i,l}} \; ,
			\label{eq:stat_combination_weight}
		\end{equation}
		where the normalization constants $\mathsf B_i$ are found using \equ{nucleon_conservation}. The 
		expression chosen for $\beta_i$ is commonly used for nuclei~\cite{Cole2010} 
		\begin{equation}
			\beta = \sqrt{\frac{A}{8 \bar \epsilon^*}} \; ,
		\end{equation}
		where the excitation energy $\bar \epsilon^*$ was chosen as the average photon energy in the 
		range where the spallation formula was obtained
		\begin{equation}
			\bar \epsilon^* = \int\limits_{0.1\;{\rm GeV}}\limits^{1\;{\rm GeV}}
			\sigma_j(\epsilon_r)\epsilon_r d\epsilon_r  \bigg / 
			\int\limits_{0.1\;{\rm GeV}}\limits^{1\;{\rm GeV}} \sigma_j(\epsilon_r) d\epsilon_r \; .
		\end{equation}
		

	\end{appendix}


\providecommand{\href}[2]{#2}\begingroup\raggedright
    
  \endgroup

\end{document}